\documentclass[aps,twocolumn,floats,pre,nofootinbib]{revtex4-1}

\usepackage{graphics,graphicx}
\usepackage{amssymb,color}
\usepackage{wrapfig}
\usepackage{amsmath}
\usepackage{amsmath}
\usepackage{amssymb}
\usepackage{graphicx}
\usepackage{physics}
\usepackage[colorlinks=true, allcolors=blue]{hyperref}
\usepackage{bm}
\usepackage{xr}
\usepackage{scrextend}

\usepackage[export]{adjustbox}

\def\eqref#1{Eq.~\ref{#1}}

\newcommand\blfootnote[1]{%
	\begingroup
	\renewcommand\thefootnote{}\footnote{#1}%
	\addtocounter{footnote}{-1}%
	\endgroup
}

\begin{document}

\title{Renormalization group study of marginal ferromagnetism}
	
\author{Andrea Cavagna}
\affiliation{Istituto Sistemi Complessi, Consiglio Nazionale delle Ricerche, UOS Sapienza, 00185 Rome, Italy}
\affiliation{Dipartimento di Fisica, Universit\`a\ Sapienza, 00185 Rome, Italy}
\affiliation{INFN, Unit\`a di Roma 1, 00185 Rome, Italy}

\author{Antonio Culla$^*$}
\affiliation{Istituto Sistemi Complessi, Consiglio Nazionale delle Ricerche, UOS Sapienza, 00185 Rome, Italy}
\blfootnote{$^*$ Corresponding author: antonio.culla@sapienza.isc.cnr.it}

\author{Tom\'as S. Grigera}
	
\affiliation{Instituto de F\'\i sica de L\'\i quidos y Sistemas Biol\'ogicos (IFLySiB) --- CONICET and Universidad Nacional de La Plata, Calle 59 n 789, B1900BTE  La Plata, Argentina}
\affiliation{CCT CONICET La Plata, Consejo Nacional de Investigaciones Cient\'\i ficas y T\'ecnicas, Argentina}
\affiliation{ Departamento de F\'\i sica, Facultad de Ciencias Exactas, Universidad Nacional de La Plata, Argentina}
\affiliation{Istituto Sistemi Complessi, Consiglio Nazionale delle Ricerche, UOS Sapienza, 00185 Rome, Italy}
	
\begin{abstract}

When studying the collective motion of biological groups a useful theoretical framework is that of ferromagnetic systems, in which the alignment interactions are a surrogate of the effective imitation among the  individuals.  In this context, the experimental discovery of scale-free correlations of speed fluctuations in starling flocks poses a challenge to the common statistical physics wisdom, as in the ordered phase of standard ferromagnetic models with $\mathrm{O}(n)$ symmetry, the modulus of the order parameter has finite correlation length. To make sense of this anomaly a novel ferromagnetic theory has been proposed, where the bare confining potential  has zero second derivative (i.e.\ it is marginal) along the modulus of the order parameter. The marginal model exhibits a zero-temperature critical point, where the modulus correlation length diverges, hence allowing to boost both correlation and collective order by simply reducing the temperature. Here, we derive an effective field theory describing the marginal model close to the $T=0$ critical point and calculate the renormalization group equations at one loop within a momentum shell approach. We discover a non-trivial scenario, as the cubic and quartic vertices do not vanish in the infrared limit, while the coupling constants effectively regulating the exponents $\nu$ and $\eta$ have upper critical dimension $d_c=2$, so that in three dimensions the critical exponents acquire their free values,  $\nu=1/2$ and $\eta=0$. This theoretical scenario is verified by a Monte Carlo study of the modulus susceptibility in three dimensions, where the standard finite-size scaling relations have to be adapted to the case of $d>d_c$. The numerical data fully confirm our theoretical results.

\end{abstract}
	
\maketitle
	
\section{Introduction}

Ferromagnetic models have been the staple of the statistical physicists' way to study collective motion in biological systems, and more generally in active matter. The seminal Vicsek model of flocking \cite{vicsek+al_95} is essentially a ferromagnetic $\mathrm{O}(n)$ model on the move, where each particle aligns its orientation to the local neighbours, but instead of being anchored on a lattice, it actively moves following its own direction. The corresponding continuous theory formulated by Toner and Tu \cite{toner+al_95, toner1998flocks, tu1998sound}, is essentially Navier-Stokes hydrodynamics meeting the Landau-Ginzburg theory of critical phenomena. Beyond these key cases, models and theories where local effective alignment plus active motion are the key ingredients, have been used across many alleys of active matter \cite{marchetti_review}. Of course, in most active systems off-equilibrium effects play a fundamental role in giving a phenomenology different from the standard framework of statistical physics; among the many examples, a very vivid one is the emergence of long-range order in the low temperature phase of the Vicsek model even in two dimensions, due to the off-equilibrium coupling between polarization and density, which propagates order more effectively than in the equilibrium case, hence bypassing the Mermin-Wagner impossibility to have ferromagnetic order in $d=2$ \cite{toner+al_95}.

In some other cases, though, the deviations of active systems from standard ferromagnetic phenomenology seem not principally due to off-equilibrium effects. In the case of biological systems this is hardly a surprise, given that being out of equilibrium is but one of the many new hurdles that biology puts in front of us when modelling living systems. The case of bird flocks is interesting, from this point of view. Experiments have shown that connected correlations are scale-free in starling flocks in the wild \cite{cavagna+al_10}. Flocks are highly ordered systems, hence in the ferromagnetic context it is reasonable to model them as (active) $\mathrm{O}(n)$ systems in their low temperature phase (which is essentially what Toner-Tu theory does), where the Goldstone theorem \cite{goldstone1961field} grants massless transverse modes, giving scale-free correlations of the orientations fluctuations. The problem, however, is that starling flocks display long range correlation also of the {\it speed} fluctuations, namely of the modulus of the order parameter. This is an anomaly in standard equilibrium systems: while the longitudinal fluctuations {(i.e. the fluctuations that, in a Cartesian orthogonal decomposition, are parallel to the total magnetization)}, which are massive at the bare level, become in fact massless after renormalization due to the coupling with the transverse modes \cite{patashinskii1973longitudinal, brezin1973feynman, brezin1973critical}, the modulus is {\it always} a massive mode in the ordered phase, and it therefore has finite correlation length. Moreover, the off-equilibrium nature of flocks does not seem to play a crucial role in connection to this anomaly, as both off-equilibrium simulations of self-propelled particles ruled by standard $\mathrm{O}(n)$ ferromagnetism \cite{hemelrijk2015scale}, and the relative theoretical approaches \cite{kyriakopoulos2016leading}, find that the speed is not a scale-free variable in the active case. This is probably not surprising, as experiments show that starling flocks are quasi-equilibrium systems, since ---due to the strong ordering--- the reshuffling time of the interaction network is significantly larger than the local relaxation time of the velocity \cite{mora2016local}. This does not exclude that off-equilibrium effects may emerge when studying these systems on very long time scales, but this would not explain the scale-free behaviour of speed fluctuations. Summing up, speed scale-free correlations are an anomaly that statistical physics should explain with some new ingredients unrelated to off-equilibrium effects.

The first attempt to explain scale-free speed correlations was done in \cite{bialek+al_14}, where a maximum entropy model derived directly from the experimental correlation data in flocks found that a standard $\mathrm{O}(n)$ ferromagnetic potential confining the modulus of the velocity can give scale-free speed correlations provided that the amplitude $g$ of the potential is small enough: within a spin-wave expansion (which holds quite well in the ordered phase of flocks), the modulus correlation length scales as $g^{-1/2}$, and because flocks are large but finite systems of linear size $L$, if $g \ll L^{-2}$, one finds scale-free speed correlations over all observable scales \cite{bialek+al_14}. The idea of this approach is to reduce the amplitude $g$ of the whole bare potential, hence reducing its curvature in the modulus direction, so to boost the correlation length beyond the system's size $L$; but because flocks are finite, this does not require $g$ to be strictly zero, hence a speed-confining potential bounding the theory is always present in the effective Hamiltonian. This promising theoretical model, however, did not stand in front of new generation of experimental data, which showed that a comparison between theory and data crashes at low values of the flocks' size $L$ \cite{cavagna2022marginal}: in small groups, the low value of the potential amplitude $g$ blows the group speed to values that far exceed the natural reference speed, and ---most importantly--- disagree with experimental observations. Essentially, what happens is that by lowering the  amplitude $g$ of the whole confining potential, we are not only decreasing the speed mass (hence increasing its correlation length), but we are at the same time depressing the bounding capacity of the potential, hence allowing the entropy to blow the collective speed to unrealistic values, which are indeed completely absent in the experimental data.

A different approach - still based on ferromagnetism - was proposed in \cite{cavagna2019CRP}, and successfully tested against numerical simulations and  - most importantly - experimental data in \cite{cavagna2022marginal}. The idea of the new theory is to have zero curvature of the bare potential from the outset, without the need to decrease the overall amplitude of the bounding potential. This can be done by switching from the classic $\mathrm{O}(n)$ bare potential, $V= g \left(1-\bm{\sigma} \cdot \bm{\sigma}\right)^2$, which bounds around one the modulus of the fluctuating variable $\bm{\sigma}$ and which needs a small $g$ to decrease the second derivative along the modulus, to the equally simple form $V=\lambda \left(1-\bm{\sigma} \cdot \bm{\sigma}\right)^4$, which has zero second derivative of the modulus {\it irrespective} of the value of the amplitude $\lambda$; because of this always-vanishing curvature, this was called {\it marginal potential} \cite{cavagna2019CRP}. The fact that the bare mass of the modulus is zero, suggests that the modulus correlations are scale-free (even in the bulk) exactly at $T=0$, where entropic effects are not present; on the other hand, upon raising the temperature, fluctuations create a non-zero curvature (that is a mass, in field-theoretical language), which decreases the modulus correlation length. A mean-field analysis showed that this is indeed the case \cite{cavagna2019CRP}: the marginal model has a finite-temperature phenomenology completely analogous to its $\mathrm{O}(n)$ cousin, with a standard ordering transition at a finite $T_c$, but it also has a new zero-temperature critical point where the modulus correlation length diverges as $\xi \sim T^{-1/2}$. Hence, in the marginal model, in order to obtain scale-free correlations in systems of finite size $L$, one simply has to push the system deeply in the ordered phase and satisfy $T \ll L^{-2}$, while the fact that the amplitude $\lambda$ is no longer connected to the modulus correlation means that it can remain finite, hence allowing the bounding potential to tame  the collective speed of the group. Results of self-propelled particle simulations ruled by the marginal confining potential are completely compatible with both the theoretical expectations and the experimental data \cite{cavagna2022marginal}, hence the marginal theory of speed control is at the moment a reasonable hypothesis to explain scale-free speed correlations in flocks.

The analytic study of the marginal theory has been limited up to now to the equilibrium mean-field approximation \cite{cavagna2019CRP}. Hence, to do theoretical progress one should first go beyond mean-field, performing a finite-dimensional study still at equilibrium, and finally extend the analysis beyond the equilibrium case, eventually including self-propulsion terms in the equations of motion. Here, we deal with the first part of this program, by writing an effective field theory for the marginal model valid in the deeply ordered phase where flocks live, namely in the vicinity of the zero temperature critical point, and by calculating the critical exponents using the Renormalization Group (RG) in momentum shell \cite{wilson1972critical, wilson1974renormalization} at one loop. Apart from the solid methodological motivation that it is better to first have a complete theoretical grasp of the equilibrium case before moving to off-equilibrium, the equilibrium theory has some interest {\it per se}. As we have already said, starling flocks are close to equilibrium, hence the equilibrium theory has great interest, if nothing else as a reference theory around which developing a future framework for small deviations from equilibrium. Finally, marginal ferromagnetism has an interesting zero temperature critical point, which is unusual under many respects even in the context of equilibrium statistical physics. The strange mix that we will find of free critical exponents and interacting theory, with relevant non-Gaussian couplings, will confirm {\it a posteriori} that the marginal theory has some intrinsic theoretical interest.

\section{The Marginal  Ferromagnetic Theory}

\subsection{Microscopic model}

The microscopic Hamiltonian of the general ferromagnetic class of models we study is given by,
\begin{equation}
  H=\frac{J}{2} \sum_{i,j}^{N} n_{ij} \left(\bm{\sigma}_i - \bm{\sigma}_j\right)^2 + \sum_{i}^N V\left(\bm{\sigma}_i \cdot \bm{\sigma}_i\right) \ , \label{erboss}
\end{equation}
where the $\bm{\sigma}_i$ are (classical) spins with $n$ components, living in an external space of $d$ dimensions. The first ferromagnetic term represents mutual imitation, favouring the spins to have similar orientation and modulus. In the finite-dimensional case, the adjacency matrix is given by $n_{ij}=1$ if $i$ and $j$ are nearest neighbours, and $n_{ij}=0$ otherwise;  $N$ is the total number of spins in the system. Spins are soft real variables, i.e.\ their modulus is not fixed, hence the bare potential $V$ has the role to bound the modulus of the spins around a reference value, which we will fix to $1$. This requirement, together with rotational invariance and the need to have a maximum at $\sigma=0$, fixes the general form of the bare potential, $V\sim \left(1-\bm{\sigma} \cdot \bm{\sigma}\right)^p$.
 The case of normal ferromagnets is given by the $p=2$  standard $\mathrm{O}(n)$ potential, $V = g  \left(1-\bm{\sigma} \cdot \bm{\sigma}\right)^2$, whose coarse-grained field theory gives the classic Landau-Ginzburg Hamiltonian \cite{parisi_book}; this theory has non-zero bare mass of the modulus, proportional to $g$, hence the correlation function of the modulus (i.e. speed correlations, in the biological context) are not scale-free in the low temperature phase, unless $g$ itself becomes small, which has its own shortcomings, as we discussed in the Introduction and demonstrated in \cite{cavagna2022marginal}.
 
The marginal model, on the other hand, is given by the $p=4$ case, namely by the following bare potential  \cite{cavagna2019CRP, cavagna2022marginal}, 
\begin{equation}
	V(\bm{\sigma} \cdot \bm{\sigma})=\lambda \left(1-\bm{\sigma} \cdot \bm{\sigma}\right)^4, \label{vice}
\end{equation}
where $\lambda$ is an amplitude. The marginal form is the simplest one with a flat minimum also in the longitudinal direction, i.e.\ a minimum with zero curvature.  With this potential, the modulus mode becomes massless at zero temperature,  irrespective of the value of $\lambda$ \cite{cavagna2019CRP, cavagna2022marginal}, hence developing scale-free correlations.  We want to investigate this zero-temperature critical point with the renormalization group \cite{wilson1974renormalization}.

\subsection{From the mean-field case to field theory}

The first step in our study is to define a field-theory version of the marginal model, to which we can then apply the momentum-shell RG method. To do this we will proceed in a phenomenological way, similar to the Landau-Ginzburg case, namely we will look for the coarse-grained field theory whose Landau approximation gives the same results as the mean-field approximation of the microscopic model \cite{parisi_book}. The mean-field theory of the marginal model was studied in \cite{cavagna2019CRP}: by setting the adjacency matrix to $n_{ij} = 1/N$ for all pairs, one obtains a fully-connected (or infinite dimensional) model where the saddle point method can be used to calculate in the limit of $N \rightarrow \infty$ the partition function of the system. If we define the magnetization as, 
\begin{equation}
\bm{m} = \frac{1}{N} \sum_i \bm{\sigma}_i \ ,
\end{equation}
and its modulus $m=|\bm{m}|$, the probability distribution of $m$ defines the mean-field Gibbs free-energy $g(m)$, 
\begin{equation}
P(m) \sim e^{-Ng(m)} \ .
\end{equation}
Working at  $T \ll 1$ and expanding $g(m)$ near $m = 1$ (which is the equilibrium magnetization at $T=0$), we obtain
(see Appendix \ref{app:mean-field} for details),	
	\begin{align}
		g(m) = &\lambda \left(1-m^2\right)^4 + \nonumber\\
		+ &T \left[a_2\left(1-m^2\right)^2+a_3\left(1-m^2\right)^3+...\right] + \nonumber \\
		+ &T^2 \left[a_1\left(1-m^2\right) + a_4\left(1-m^2\right)^2+...\right], \label{gigi}
	\end{align}
where the $a_n$ are $T$-independent constants which are functions of the parameters $J$ and $\lambda$ of the Hamiltonian.  For $T=0$ the free energy reduces to the same functional form as $V$, \eqref{vice}, and it thus has a minimum with zero curvature.  So the mean-field Gibbs free energy, in the limit of vanishing temperature, has a flat minimum, implying a divergent susceptibility for fluctuations of the modulus of the magnetization.  On the other hand, when $T$ grows, entropic fluctuations generate a non-zero second derivative of the free energy, hence making the susceptibility finite. This trade-off between bare potential and entropic fluctuations close to $T=0$ is the origin of the zero-temperature critical point of the marginal model. This mean-field scenario was confirmed also in the finite-dimensional case by numerical simulations on a cubic lattice \cite{cavagna2019CRP}.

We can reorder the terms in \eqref{gigi}, collecting powers of $(1-m^2)$ and writing the coefficients to the lowest order in $T$,
\begin{multline}
	g(m) = a_1 T^2 \left(1-m^2\right) + a_2 T \left(1-m^2\right)^2 + \\
	+ a_3 T\left(1-m^2\right)^3 + \lambda \left(1-m^2\right)^4 + \ldots \label{latrottola}
\end{multline}
To proceed in defining the field theory, we do not need the actual values of the coefficient $a_n$, as the only relevant thing is that they do not depend on the temperature $T$.  We now promote  the magnetization {modulus} to a fluctuating field, $m \rightarrow \phi(\bm{x})$.  Because we are interested in the system's properties near the marginal critical point \cite{cavagna2019CRP} at $T=0$, where the equilibrium magnetization modulus is $1$, it is convenient to work with the shifted field, $\varphi(\bm{x}) = 1 - \phi(\bm{x})$, which is small near the zero-temperature critical point. {We stress the fact that, even if the magnetization modulus is not analytic for $m$ close to $0$, we are far from this regime since in the low temperature phase $m\simeq 1$. Following this scheme we have that $(1-m^2)=(1-m)(1+m) \rightarrow 2\varphi$, where the numerical factor $2$ will be absorbed into the couplings of the field theory.}  Additionally, we ignore the angular degrees of freedom, focusing only on the modulus fluctuations, because the fluctuations of modulus and phase are known to be very weakly coupled to each other in the broken-symmetry phase  \cite{brezin1973critical, patashinskii_book,ryder1996quantum}.  Finally, following the standard ferromagnetic procedure, we introduce a square gradient term, which embodies ferromagnetic interaction by depressing short-wavelength fluctuations of the field. By keeping powers up to $\varphi^4$ (higher order terms are discussed in Appendix C), we finally obtain the following Landau free-energy,
\begin{equation}
  \mathcal{F} = \int \mathrm{d}^d x \left\{ \left(\nabla \varphi\right)^2 + aT \varphi^2 + cT^2 \varphi + vT \varphi^3 + u \varphi^4 \right\} \ , \label{newbie}
\end{equation}
so that the probability of a field configuration is, $P[\varphi] = \exp[-\mathcal{F}/T] /\mathcal{Z}$. In conventional field theories \cite{goldenfeld_lectures_1992} we normally would ignore the factor $1/T$ in the exponential weight, because near the critical point it contributes a harmless finite constant $1/T_c$ that can be safely reabsorbed in the field and in the couplings.  In our case, however, we must be careful, as we are dealing with a critical point at $T_c=0$, hence $T$ is {\it not} a harmless constant.  The temperature is the coefficient of the quadratic term, and it therefore plays the role of the bare mass; however, note that powers of $T$ appear also in the other coefficients, not just in the quadratic one, so that when approaching the critical temperature, all these coefficients vanish. For this reason one cannot reabsorb the temperature in the other couplings.
The most convenient way to deal with this situation is to define a new field,
\begin{equation}
 \psi(\bm{x}) =  \varphi(\bm{x})/ \sqrt{T}  \ .
\end{equation}
This rescaling leads to a theory with a regular coefficient of the square gradient term, and results in a field amplitude that does not vanish for $T \rightarrow 0$ (see App.~\ref{app:field-rescaling}).
We will also drop the linear term, which does not change the critical behavior of the theory (this is justified in App.~\ref{sec:linear-term}),
and set the constant $a$ to $1$, which amounts to a harmless redefinition of the temperature and of the other couplings.  We thus end up with a Landau-Guinzburg theory for $\psi(x)$ such that $P[\psi]=\exp[-\mathcal{H}]/ \mathcal{Z}$, with,
\begin{equation}
		\mathcal{H} = \int \mathrm{d}^d x \left\{ \left(\nabla \psi\right)^2 + T \psi^2 + vT^{3/2} \psi^3 + uT \psi^4 \right\} \ .
		 \label{Jimi}
	\end{equation}
The novelty of this field theory is that powers of $T$, which here plays the role of the mass (i.e. of the control parameter), appear in all the couplings.  This is unusual in standard field theories, where the bare couplings are independent of the temperature (or mass), and thus remain finite when the bare mass vanishes.
	
Dimensional analysis of \eqref{Jimi} shows that the naive scaling dimensions are (in momentum units),
\begin{align}
		[k] & = 1  & [\psi_k]& =-\frac{d}{2}-1  &  [T]&=2 \nonumber\\
		[v]&=-\frac{d}{2}  &  [u]&=2-d , \label{naive}
\end{align}
where $\psi_k$ is the field in momentum space.  We immediately see that for $d>2$ the naive scaling dimensions of both $v$ and $u$ are negative, suggesting that for $d=3$ the theory is infrared-free. However, computing the naive dimensions of the {\it full} cubic and quartic coefficients, $vT^{3/2}$ and $uT$, we find
\begin{align}
		[vT^{3/2}] &=\frac{6-d}{2} &  [uT]&=4-d,
\end{align}
so that for $d=3$ their naive scaling dimension is positive, suggesting therefore that the theory actually conserves its non-Gaussian couplings in the infrared limit, so that it is {\it not} free.  This apparently contradictory situation needs to be settled by going beyond mere dimensional analysis, that is by calculating the renormalization group flow equations.

\section{Renormalization Group analysis}

\subsection{General RG procedure}
	
We study the zero-temperature critical behaviour of the Hamiltonian \eqref{Jimi} using Wilson's momentum-shell renormalization group method \cite{wilson1971renormalization1}.  We present in this section the recursion relations of the RG transformation. The diagrammatic perturbation theory can be carried out using the tuning parameter $T$, and the two composite coupling constants, 
\begin{equation}
\hat v= vT^{3/2} \quad , \quad  \hat u=uT.
\end{equation}
Formally, then, all diagrams are the same as in the standard Landau-Guinzburg theory (with a cubic term). However, after having worked out the RG flow equations for $(T, \hat v, \hat u)$, it will be crucial to go back and study the RG flow of the \emph{original} parameters $(T, v, u)$ to understand the critical behavior, which is different from that of standard Landau-Guinzburg in $d=3$. In fact, neglecting the explicit $T$-dependence of $\hat v$ and $\hat u$ leads to physical inconsistencies that are already apparent at the level of the Landau approximation: if one looks for a constant solution, $\psi(x)=\psi_0$ (thus setting to zero the gradient square) and simply minimizes $\mathcal{H}$ with respect to $\psi_0$, one finds that for fixed $\hat v$ and $\hat u$ the potential has {\it two} minima, one at $\psi_0=0$ and second one at finite value of $\psi_0$ with {\it lower} energy, giving a first-order transition phenomenology.  Instead, working with $T\to0$ at fixed $v$ and $u$ keeps the appropriate balance among the coefficients such that the Landau potential always has just {\it one} minimum at $\psi_0=0$, which is consistent with the mean-field scenario.

To do momentum-shell RG one first rewrites the Hamiltonian in momentum space, introducing an arbitrary ultraviolet cut-off $\Lambda$ (of the order of the inverse of the nearest-neighbor distance) that makes all perturbative diagrams well-behaved in the UV limit.
The Wilson procedure then consists of two steps \cite{wilson1974renormalization}.  First one integrates out all the degrees of freedom in a thin shell $k\in [\Lambda/b,\Lambda]$ with $b>1$ but close to 1, defining,
\begin{equation}
  \mathcal{H}_1[\psi(k\le \Lambda/b)] \equiv -\log \int\!\!D[\psi(\Lambda/b<k\le \Lambda)] e^{-H[\psi]}.
\end{equation}
This step is non-trivial because the non-Gaussian terms couple the on-shell (UV) and off-shell (IR) modes, and must be carried out perturbatively.  Once $\mathcal{H}_1$ is found, in the second step the momentum is rescaled $k\to k/b$ so that the original cut-off is recovered, and the coarse-grained Hamiltonian is re-written so that it has the same form as the original one, but with new, renormalized field and coupling constants. As a result of the two steps we obtain the novel Landau-Ginzburg Hamiltonian, 
\begin{multline}
\mathcal{H}_{b}\left[\psi_b(\bm{k})\right] = \frac{1}{2} \int_0^{\Lambda} \!\! \frac{\mathrm{d}^d k}{(2\pi)^d}\, \left[ \left(k^2 + T_b\right) \psi_b(\bm{k}) \psi_b(-\bm{k}) \right] +  \\
+ \hat v_b\int_0^{\Lambda} \!\!\frac{\mathrm{d}^d k_1 \mathrm{d}^d k_2}{(2\pi)^{2d}}\,  \psi_b(\bm{k}_1) \psi_b(\bm{k}_2) \psi_b(-\bm{k}_1-\bm{k}_2) + \\
+ \hat u_b\int_0^{\Lambda}\frac{\mathrm{d}^d k_1 \mathrm{d}^d k_2 \mathrm{d}^d k_3}{(2\pi)^{3d}}\,  \psi_b(\bm{k}_1) \psi_b(\bm{k}_2) \psi_b(\bm{k}_3) \times \\ \times \psi_b(-\bm{k}_1-\bm{k}_2-\bm{k}_3)
\end{multline}
which depends on the renormalized couplings  $T_b, \hat v_b, \hat u_b$, and the renormalized field $\psi_b(k)$. In order to find these renormalized parameters we need to turn to the diagrammatic expansion at one loop.

\subsection{Relevant diagrams and RG relations}
	
\begin{figure}
  \centering
  \includegraphics[angle=-90,width=\columnwidth]{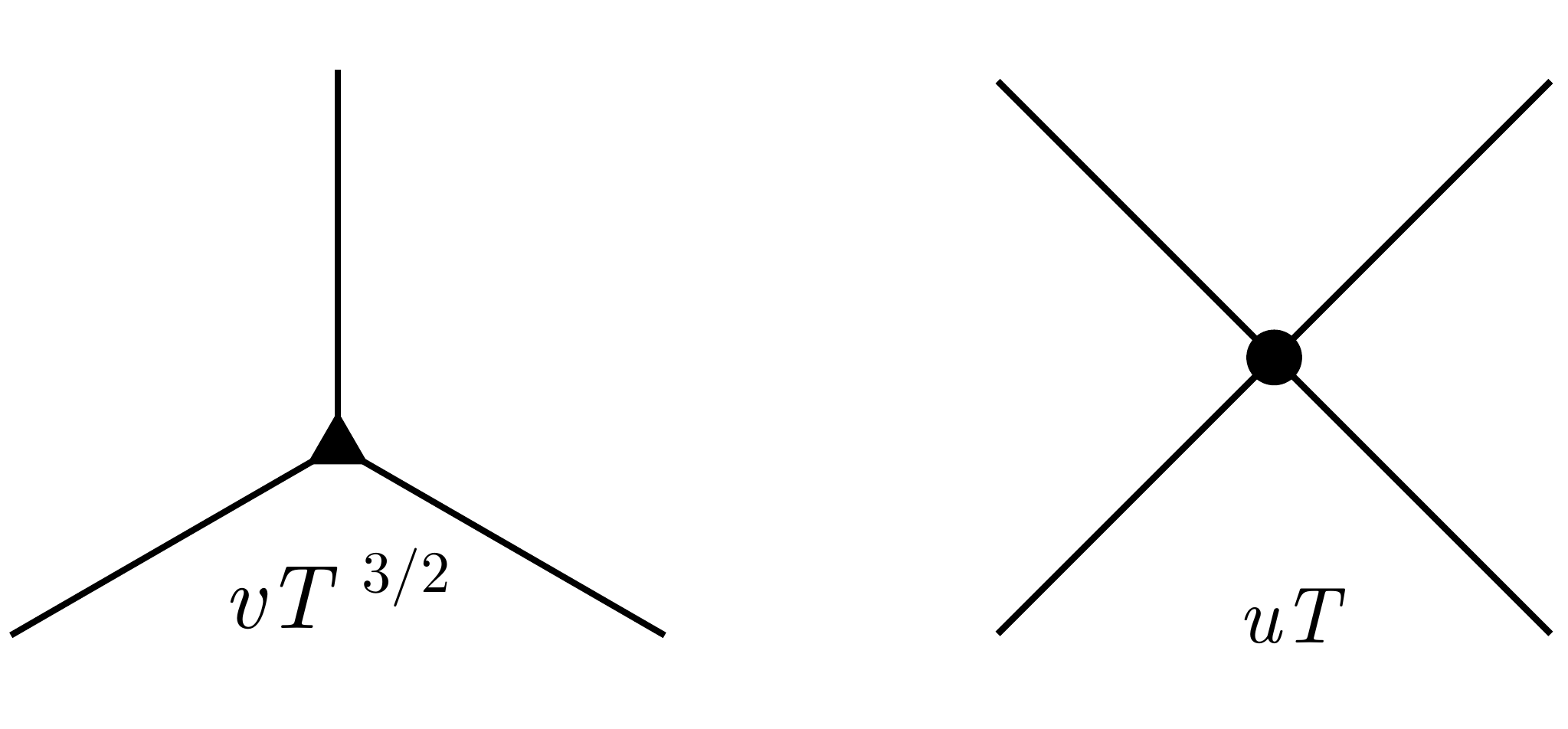}
  \caption{The two vertices of the marginal field theory.}
  \label{fig:vertices}
\end{figure}

\begin{figure}[h]
  \centering
  \includegraphics[angle=-90,width=\columnwidth]{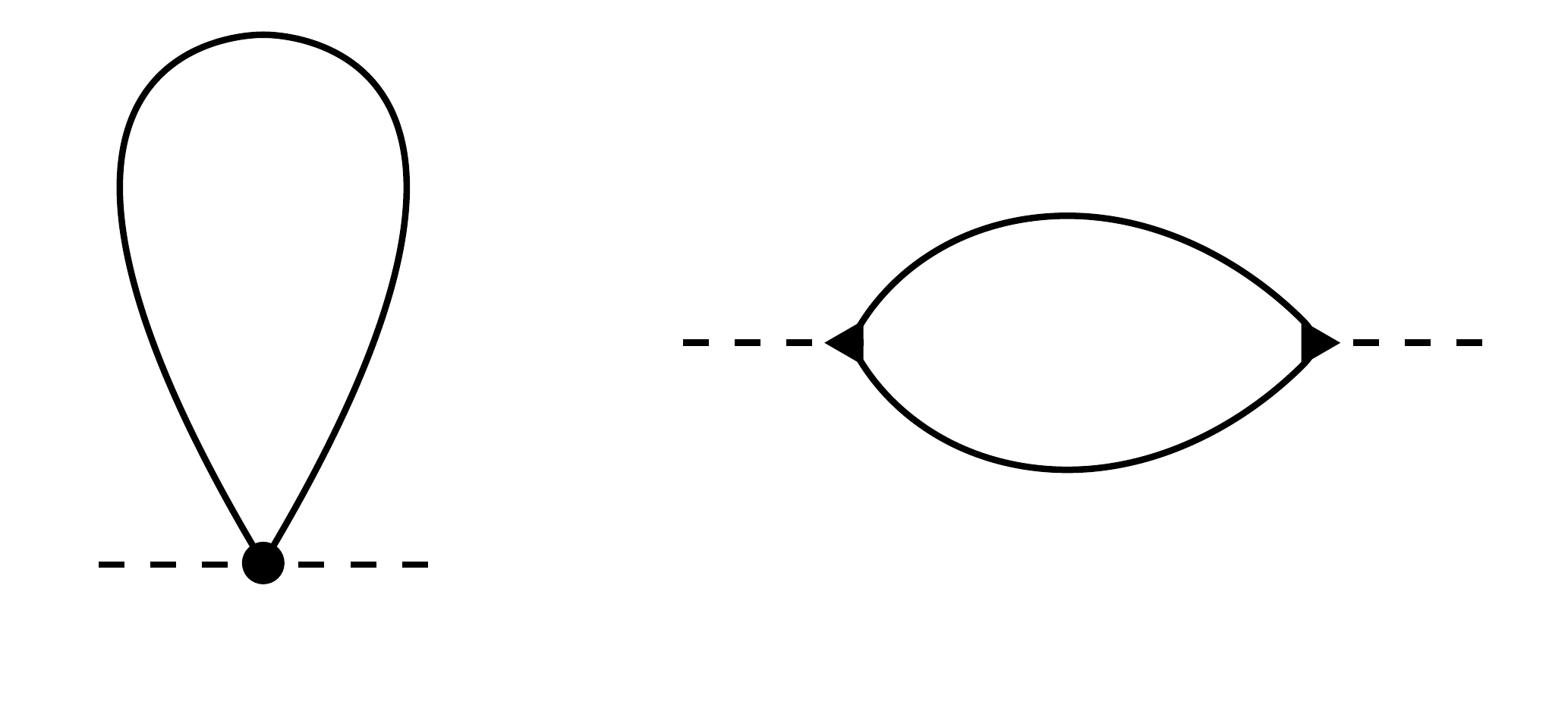}
  \caption{Diagrams that contribute to the renormalization of $T$ and field.  Dashed lines represent fields with momentum $k<\Lambda/b$ (off-shell), while solid lines represent integrated fields with on-shell momentum $\Lambda/b < k < \Lambda$.} 
  \label{fig:two-legs}
\end{figure}

\begin{figure}[h]
  \centering
  \includegraphics[angle=-90,width=\columnwidth]{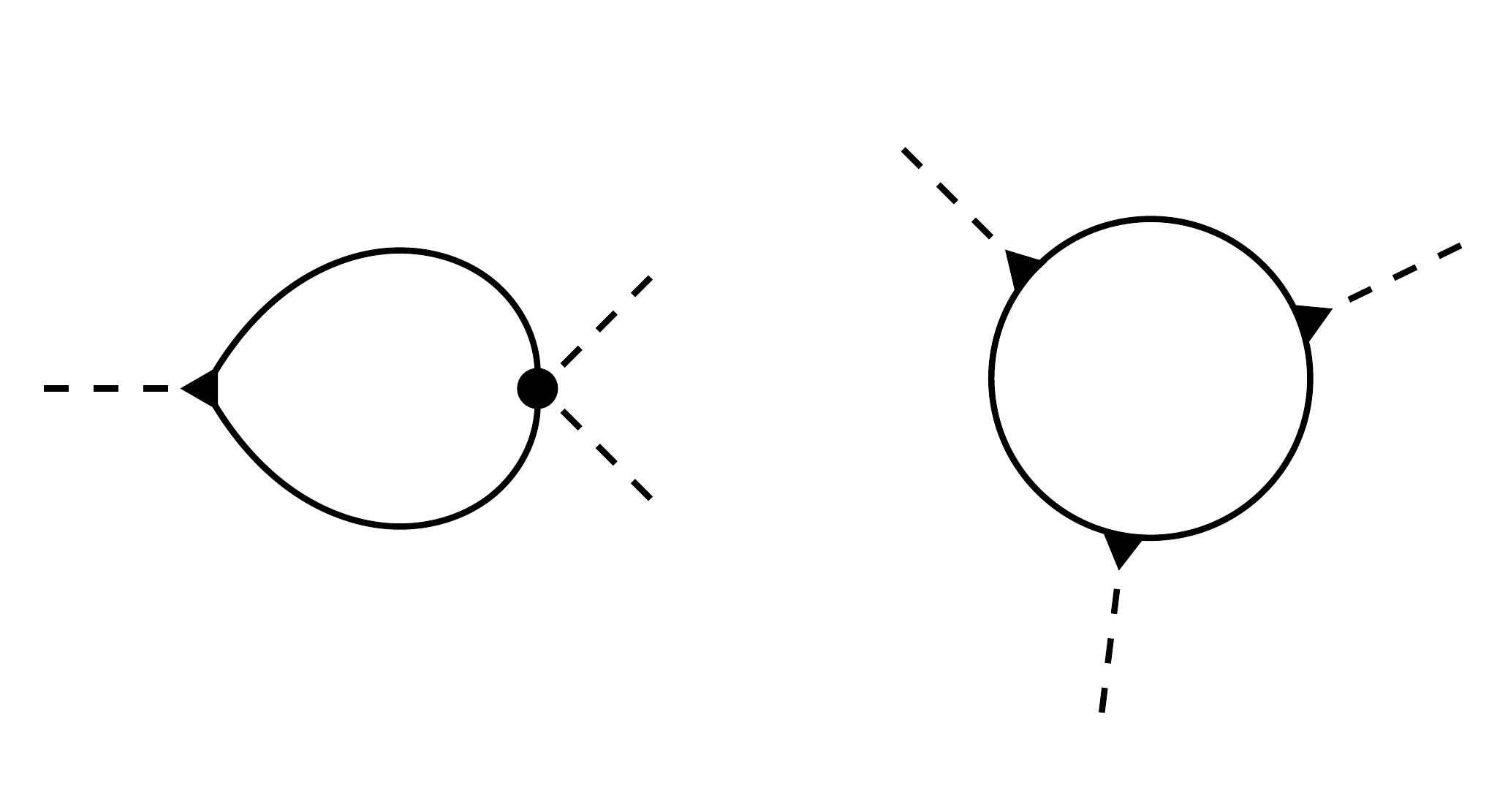}
  \caption{Diagrams contributing to the renormalization of $\hat v=vT^{3/2}$.}
  \label{fig:three-legs}

\end{figure}

\begin{figure}[h]
  \centering
  \includegraphics[angle=-90,width=\columnwidth]{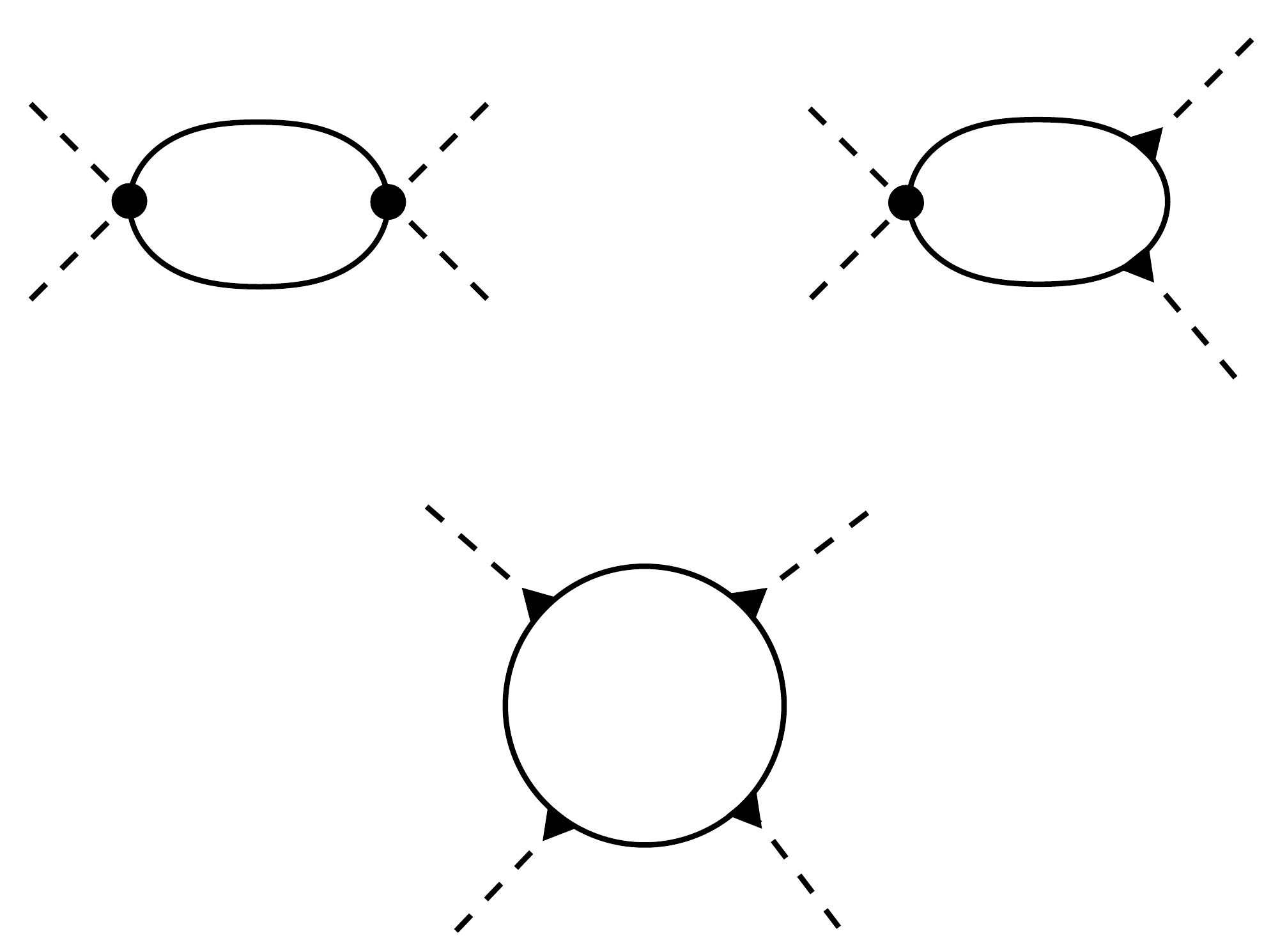}
  \caption{Diagrams contributing to the renormalization of $\hat u= uT$.}
  \label{fig:four-legs}
\end{figure}

The theory has two vertices (Fig.~\ref{fig:vertices}), a cubic one with coupling $\hat v= vT^{3/2}$ and a quartic one with coupling $\hat u= uT$.
Combining these two vertices we can make one-loop diagrams with an arbitrary number of external legs, but we evaluate the renormalized couplings only up to the $\psi^4$ term (four external legs).  Diagrams with more than four external legs give a correction to higher order terms that we do not include in Hamiltonian \eqref{Jimi} because they are all RG-irrelevant (see App.~\ref{app:RGdetails}).  We have two diagrams that contribute to the renormalization of temperature $T$ and field (Fig.~\ref{fig:two-legs}), two that enter the renormalization of $vT^{3/2}$ (Fig.~\ref{fig:three-legs}) and three that contribute to the renormalization of $uT$ (Fig.~\ref{fig:four-legs}).


Combining the contributions of all diagrams, the renormalized couplings are found to be (details in App.~\ref{app:RGdetails})
\begin{subequations}
  \begin{align}
  T_b ={}& b^2 \left[T + 12 u T A_1 - 18 v^2 T^3 A_2\right] \label{athos}\\
  \left(vT^{3/2}\right)_b ={}& b^{3-d/2} \Big[vT^{3/2} - 36 u v T^{5/2} A_2 + \nonumber \\
         & \qquad + 36 v^3 T^{9/2} A_3\Big] \label{porthos}\\
  \left(uT\right)_b ={}& b^{4-d} \Big[uT - 36 u^2 T^2 A_2 +216 u v^2 T^4 A_3 + \nonumber \\
         & \qquad - 162 v^4 T^6 A_4\Big], \label{aramis}
  \end{align}
  \label{eq:renormhat}%
\end{subequations}
where
\begin{equation}
  A_q = \int_{\Lambda/b}^{\Lambda} \frac{\mathrm{d}^d p}{(2 \pi)^d} \frac{1}{(p^2 + T)^q} \simeq K_d \frac{\Lambda^d}{(\Lambda^2 + T)^q} \log b \ , \label{pippo}
\end{equation}
{$K_d$ is the area of the unit sphere in $d$ dimensions divided by $(2\pi)^d$} and the approximation is valid for a thin shell ($b \simeq 1$).
Finally, from the $k$-dependence of the two-legged diagrams (Fig.~\ref{fig:two-legs}), the field renormalization is found as,
\begin{equation}
  \psi_b(\bm{k}) = b^{d_\psi} \psi(\bm{k}/b) \ ,
\end{equation}
where the scaling dimension of the field is given by the $k^2$ contribution of the diagram in Fig. \ref{fig:two-legs} - right (for its detailed expression see Appendix \ref{app:RGdetails}),
\begin{equation}
  d_\psi  = -1-\frac{d}{2} + \frac{B}{2} v^2T^3 \Lambda^{d-6},
           \label{eq:field-ad}
\end{equation}
and $B$ is a dimensionless numerical constant whose value we will not need in the following.

\subsection{The beta functions}

We now ``unpack'' Eqs.~\ref{eq:renormhat} to obtain the RG equations for the original coupling constants, $(T, v, u)$. Moreover, instead of
keeping the RG equations in their iterative form, we will switch to the fairly more compact differential form,  introducing the standard $\beta$-functions for each coupling  \cite{goldenfeld_lectures_1992}. To do this, one defines the infinitesimal parameter $x \ll 1$, such that $b=1+x$ and $\log b \approx x$; in this way the $\beta$-function (or flow function) of a generic parameter $\cal P$ is defined as, 
\begin{equation}
\beta_{\cal{P}}=\partial {\cal P}/\partial (\log b) =\partial {\cal P}/\partial x \ .
\end{equation} 
After using ~\eqref{eq:renormhat} to work out the flow of the original couplings, their $\beta$-functions become, 
\begin{subequations}
  \begin{align}
    \beta_T &= 2T + 12 uT K_d \Lambda^{d-2} -12uT^2 K_d \Lambda^{d-4} , \\
    \beta_v &= - \frac{d}{2} v - 18 uv K_d \Lambda^{d-2} + 18 uvT K_d \Lambda^{d-4} \nonumber \\
    		&- 36 uvT K_d \Lambda^{d-4}, \\
    \beta_u &= (2-d) u - 12 u^2 K_d \Lambda^{d-2} + 12 u^2 T K_d \Lambda^{d-4} \nonumber \\
    		&- 36 u^2 T K_d \Lambda^{d-4},
  \end{align}
  \label{eq:betafuns}%
\end{subequations}
{where we have written only the leading term and the first correction in $T$.}

\section{Fixed point and critical exponents}

From the zeros of the $\beta$ functions \eqref{eq:betafuns} we find that the RG flow has only one physically meaningful (i.e.\ with $T \geq 0$ and $u \geq 0$) fixed point, namely,
 \begin{equation}
 T^*=0   \quad , \quad v^*=u^*=0 \ .
\end{equation}
The Jacobian matrix at this fixed point is
\begin{equation}
  \frac{\partial(\beta_T,\beta_v,\beta_u)}{\partial(T,v,u)} =
  \begin{pmatrix}
    2 & 0 & 0\\
    0& -d/2&0\\
    0 & 0 & 2-d
  \end{pmatrix}
  \equiv
  \begin{pmatrix}
    y_T & 0 & 0\\
    0& y_v & 0 \\
    0 & 0 & y_u
  \end{pmatrix},
\end{equation}
from which we see that $T$ is an unstable direction, as expected, given that $T$ is the tuning parameter, while both $u$ and $v$ are stable in $d=3$.  The critical manifold is the $T=0$ plane, and $T$ is the (relevant) control variable that takes the system away from the critical point.  The critical point is $T_c=0$, independently of the (bare) value of $u$ and $v$, and independently of the cutoff $\Lambda$. Notice that, consistently with the physics of the problem, there is no negative shift of the mass, as there is instead in the standard Landau-Ginzburg theory \cite{goldenfeld_lectures_1992}: the zero temperature bare critical point cannot be reduced further by fluctuations under renormalization.

\subsection{Critical exponents}

Critical exponents can be found as usual from the eigenvalues of the Jacobian, once we linearize the RG transformation near the fixed point \cite{cardy1996scaling}. In particular, to calculate the exponent $\nu$, defining the divergence of the modulus correlation length, 
\begin{equation}
\xi \sim T^{-\nu}
\end{equation}
we use the fact that $\xi_b=\xi/b$, which gives, $\partial \xi/\partial x= -\xi$ (the correlation length has always scaling dimension $-1$), so that $\nu^{-1}$ is the scaling dimension of the control parameter, namely it is the coefficient of the linear term $T$ in the $\beta$-function of the temperature,
\begin{equation}
\nu^{-1} = \left. \frac{\partial \beta_T}{\partial T}\right|_{u^*,v^*} =  2+ 12 u^* K_d \Lambda^{d-2}= 2 \ ,
\end{equation}
where we have used the fixed point value, $u^*=0$.
We conclude that the divergence of the modulus correlation length is ruled by the same critical exponent as the free theory, $\nu=1/2$.
It is important to note that this result is due to the fact that the coefficient of the linear  term $T$ in the $\beta$-function of the control parameter depends on $u$ and {\it not} on $\hat u$. This is the reason why the exponent is free, even though the effective coupling $\hat u= Tu$ is not asymptotically zero. Notice that, had we kept hidden into $\hat u$ the dependence on the temperature in the function $\beta_T$, we would have found a fixed point at a {\it negative} value of $T$, which is clearly unphysical.

The second exponent we are interested in is the anomalous dimension of the space correlation function, $\eta$, defined by its scaling form near the critical point \cite{cardy1996scaling}, 
\begin{equation}
C(k) = k^{-2+\eta}f(k\xi) \ .
\end{equation}
From the renormalization of the field thorough the RG transformation we can write a self-consistency equation for the correlation function,  
\begin{equation}
\langle \psi(k)\psi(k')\rangle = \\
  b^{-2d_\psi^*} \langle \psi_b(bk)\psi_b(bk')\rangle 
\end{equation}
and by using the standard relation, $(2\pi)^d \delta(k+k')C(k) = \langle \psi(k)\psi(k')\rangle$, we obtain,
\begin{equation}
C(k) =   b^{-2d_\psi^*-d} C(bk) ,
\end{equation}
from which we can read the anomalous dimension,
\begin{equation}
\eta =  2+d+2d^*_\psi \ ,
\end{equation}
where  $d_\psi^*$ is the dimension $d_\psi$ evaluated at the fixed point.  From~\eqref{eq:field-ad} we find then,
\begin{equation}
  \eta = 2 + d -2 -d + \left. B v^2T^3 \Lambda^{-3}/2 \right|_{T^*, v^*,u^*} = 0.\label{eq:1}
\end{equation}
We conclude that both critical exponents take their free-theory values, 
\begin{equation}
\nu=1/2 \quad , \quad \eta=0 \ .
\end{equation}
It might seem surprising to obtain these values in $d=3$, where it is known that the cubic and quartic Landau-Ginzburg terms are relevant in the RG sense.  However, our result is a consequence of the peculiar way in which the quadratic, cubic and quartic coefficients are tied together in this theory.  If one goes back to look for fixed points in the composite couplings, Eqs.~\ref{eq:renormhat}, one does find a Wilson-Fisher-like fixed point, but it is nonphysical for this case because - as we have already noted - it would require $T^*<0$.  One can verify that, for any starting point $(T, \hat v,\hat u)$ with positive couplings and near $T=0$, the flow always stays in the region with $T>0$, which is evident considering the flow in $(T,u,v)$ space, where $T=0$ is the critical manifold.

{We should remark that, unlike the usual $\lambda \phi^4$ theory, here the critical exponent $\eta$ does pick up corrections at one loop, coming from the diagram built by combining  two $\psi^3$ vertices (which has two external legs and a non-zero external momentum on internal lines, see appendix figure \ref{twoleg}). However, this correction vanishes due to the Gaussian nature of the fixed point that rules the critical exponents in this case.  For this reason, higher order corrections to the anomalous dimension $\eta$ will also vanish.}

\subsection{Critical region}

The critical point of this theory is rather pathological, since at $T=0$ all but the gradient terms vanish.  Hence, we wish to understand whether there is some {\it finite} neighbourhood of the critical point where the free critical exponents calculated above can actually be observed. In other words, we must estimate the size of the critical region, i.e.\ the region outside which one expects noticeable departures from the power laws with the fixed-point values of the exponents.  To do this we need to go beyond the linear approximation of the flow near the fixed point. Hence, we go back to the $\beta$ functions~\eqref{eq:betafuns} and rewrite them keeping terms up to $O(T)$,
\begin{align}
  \frac{d T}{dx} &= \beta_T = 2 T \left(1+ 6 u  \Lambda^{d-2} \right) \nonumber ,  \\
  \frac{d v}{dx} &= \beta_v = -\frac{d}{2} v - 18 u v \Lambda^{d-2} \label{macarena},  \\
  \frac{d u}{dx} &= \beta_u  = (2-d)u - 12 u^2 \Lambda^{d-2} \nonumber,
\end{align}
where we have set $K_d=1$ to simplify the notation.
These equations can be solved exactly.  In $d=3$ we obtain,
\begin{align}
  T(x) &= T_0 (12 \Lambda u_0 + 1) e^{2x} - 12 \Lambda T_0 u_0 e^x \nonumber \\ 
  v(x) &= \frac{v_0 e^{-\frac{3}{2}x}}{\left(12\Lambda u_0 +1 - 12 \Lambda u_0 e^{-x}\right)^{3/2}} \label{sol} \\
  u(x) &= \frac{u_0 e^{-x}}{12\Lambda u_0 +1 - 12 \Lambda u_0 e^{-x}}, \nonumber
\end{align}  
where $T_0$, $v_0$ and $u_0$ are the physical (i.e. bare) values of the theory's parameters, that is the starting points, at $x=0$, of the RG transformation.

The critical power-law behaviour ruled by the RG fixed point can actually be observed only if the flow carries the irrelevant (stable) variables close enough to their fixed point while still remaining in the region of $T$ where the linear approximation is valid; therefore, to estimate bounds for the critical region we follow the flow using \eqref{sol} and check whether or not at the end of the flow the linear approximation is still valid.
We start the flow at $v_0 \sim O(1)$ and $u_0 \sim O(1)$, thus selecting a particular theory, and at some $T_0$  such that the physical correlation length is much larger than the lattice spacing, $\xi_0 \gg 1/\Lambda$.  The flow cannot be continued beyond the point where the correlation length approaches the lattice spacing, so we require $\xi(x_{stop}) \simeq 1/\Lambda$.  If we are in the critical region, then  $\xi_0 \sim T_0^{-1/2}$ so the stop condition implies, 
\begin{align}
  T_0 \sim \Lambda^2 e^{-2x_{stop}} , \qquad \text{or} \qquad
  e^{x_{stop}} \sim \Lambda T_0^{-1/2}\label{antiG} \ .
\end{align}
We now require that at $T(x_{stop}),u(x_{stop}),v(x_{stop})$  the linear approximation remains valid, which we can check by evaluating the $\beta$ functions~\eqref{macarena} and comparing them with the linear approximation.  From~\eqref{macarena} we see that this needs $u(x_{stop})\ll 1$, which inserting the value of $x_{stop}$ in \eqref{sol} gives the condition, 
\begin{equation}
  T_0 \ll u_0^{-2} \label{eq:aG1}.
\end{equation}
For the validity of the result $\eta=0$ we need that $d_\psi$ from \eqref{eq:field-ad} at $x_{stop}$ does not differ from $d_\psi^*$.  This requires $v^2(x_{stop}) T^3(x_{stop}) \ll 1$, that is, 
\begin{equation}
  T_0 \ll v_0^{-4/3}. \label{eq:aG2}
\end{equation}
Conditions \eqref{eq:aG1} and \eqref{eq:aG2} tell us that, for any reasonable value of physical couplings $v_0$,  $u_0$, we can choose a small enough ---but finite--- physical temperature $T_0$, below which the theory will be in the critical regime with free exponents. Considering that any reasonable values of the bare physical parameters will always be of order one, conditions \eqref{eq:aG1} and \eqref{eq:aG2} tell us that the theory will have a rather comfortable critical region above $T_c=0$. These calculations can be generalized for any $d>2$, hence we conclude that the marginal theory is infrared-free \cite{ryder1996quantum} with an upper critical dimension $d_c=2$.

{If we want to check if the conditions \eqref{eq:aG1} and \eqref{eq:aG2} are reasonable for actual finite-size implementations of the marginal model and compare the results with experiments we can see the work \cite{cavagna2022marginal}. With just a single set of parameters with a low enough temperature it is possible to reproduce scale-free correlations for all the experimental systems, obtaining also a magnetization (that in \cite{cavagna2022marginal} is called polarization) which is compatible with the experimental ones \cite{cavagna2022marginal}. The actual critical exponents may be influenced by non-equilibrium dynamical effects \cite{cavagna2013boundary} but the scale-free phenomenology is the same for data, Self Propelled Particle (SPP) simulations \cite{cavagna2022marginal} and the equilibrium model here presented.}

\section{Finite size scaling and numerical validation}

\begin{figure*}
	\centering
	\includegraphics[width=\textwidth]{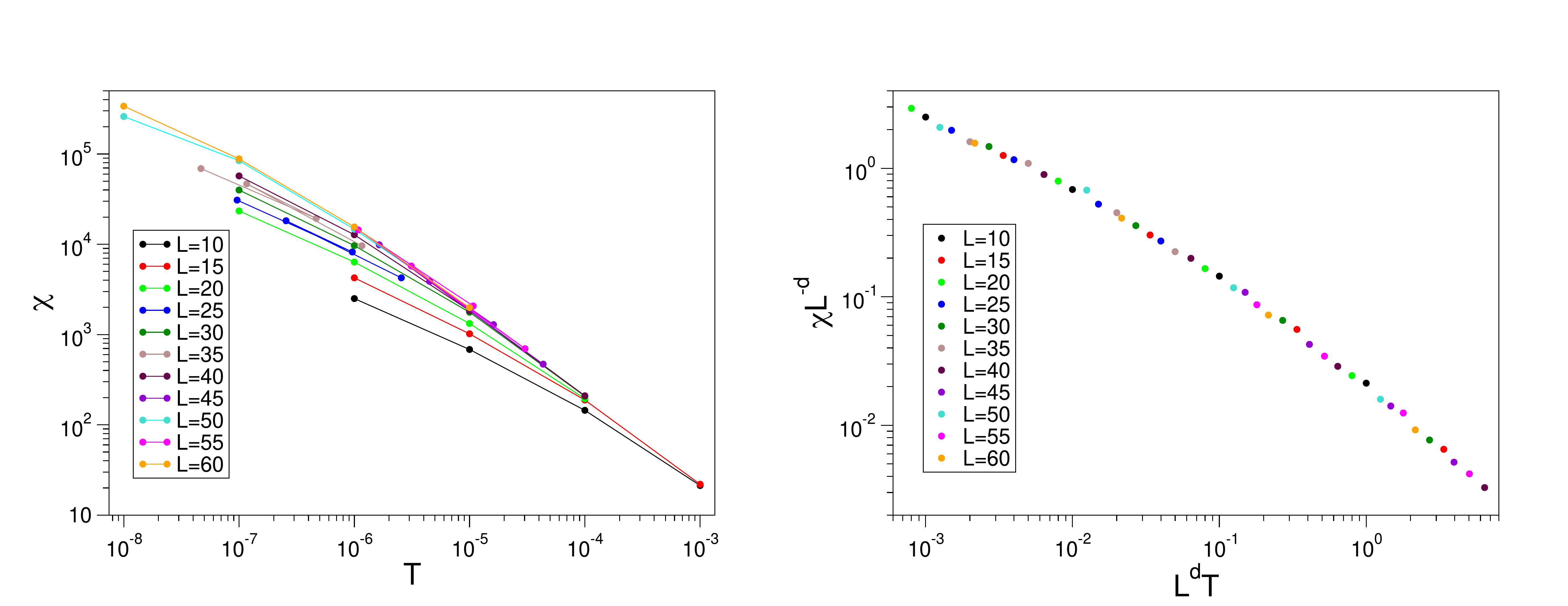}
	\caption{\textbf{Marginal model modulus susceptibility.} The modulus susceptibility of the marginal model, computed from Monte Carlo via \eqref{fdtMRG}, is shown for various sizes.   Left panel: susceptibility $\chi$ vs.\ temperature.  Right panel: rescaled susceptibility vs.\ the scaling variable $L^d T$ (\eqref{MRGscal}).  Error bars are smaller than the symbol size.  The collapse is very good, confirming that the $T=0$ critical point is infrared-free and that the marginal field theory, \eqref{Jimi} correctly describes the modulus mode of the microscopic model \eqref{erboss}. }
	\label{fig:FSS}
\end{figure*}

In order to check the validity of the theoretical calculations we resort to simulations and finite-size scaling to investigate the marginal critical point at $T=0$.  We first recall the basic results of finite-size scaling theory above the upper critical dimension, since this case is different from the more usual situation where finite-size scaling is applied, i.e.\ below the critical dimension.

For conventional ferromagnetic/paramagnetic critical points in three dimensions, the finite-size scaling for the susceptibility has the general form \cite{cardy2012finite},
\begin{equation}
\chi = L^{\gamma/\nu} f((T-T_c)L^{1/\nu}), \label{eq:fss-below}
\end{equation}
where $f(x)$ is a scaling function and $T_c$ is the critical temperature. $\gamma$ and $\nu$ are the usual critical exponents \cite{patashinskii_book}. However, since our theory is infrared-free for $d=3$, hyperscaling does not hold \cite{brezin1982investigation} and Eq.~\ref{eq:fss-below} is not valid.  To find the correct scaling we start, following \cite{brezin1985finite}, from the Landau-Ginzburg Hamiltonian \eqref{Jimi} in its Landau approximation for a finite system,
\begin{equation}
\mathcal{H} = L^d \left\{T \psi_0^2 + v T^{3/2} \psi_0^3 + u T \psi_0^4 \right\} \label{MFaction}
\end{equation}
where $\psi_0$ is a space-homogeneous field which represents the zero mode of the theory.  This amounts to neglecting diagrams with loops, which can be shown not to contribute to the scaling \cite{brezin1985finite}.  At 0 loops the susceptibility is given by,
\begin{equation}
  \chi \sim \frac{\int \mathrm{D} \psi_0 \ \psi_0^2 e^{-\mathcal{H}}}{\int \mathrm{D} \psi_0 \ e^{-\mathcal{H}}} \ . \label{chidim}
\end{equation}
Since we want to evaluate the integrals above via a saddle point it is convenient to change variable $\psi_0 \rightarrow \psi_0/(L^{d/2}T^{1/2})$ and write the action \eqref{MFaction} as
\begin{equation}
	\mathcal{H} = \psi_0^2 + \frac{v}{L^{d/2}} \psi_0^3 + \frac{u}{L^d T} \psi_0^4. 
\end{equation}
Then the susceptibility can be written as
\begin{equation}
	\chi = L^d f\left(\frac{v}{L^{d/2}},\frac{u}{L^dT}\right).
\end{equation}
For fixed $v$ and $u$, and for $L$ large enough such that we can ignore the dependence of the function $f$ on its first argument we obtain,
\begin{equation}
	\chi = L^{d} f\left(L^d T\right). \label{MRGscal}
\end{equation}
We therefore conclude that the marginal theory has an anomalous finite-size scaling behaviour due to the fact that its critical point is on the basin of attraction of an infrared-free fixed point.  In general, infrared-free theories (e.g.\ $\lambda \phi^4$ for $d>4$, which is studied for example in \cite{wittmann2014finite}) have an anomalous scaling that is usually $\chi = L^{d/2} f((T-T_c)L^{d/2})$ \cite{brezin1985finite}.  For the marginal model, however, the peculiar dependence on $T$ of the couplings leads to a different scaling form, \eqref{MRGscal}. One can include in this discussion higher order terms of the marginal field theory Hamiltonian, but it can be easily verified that their contribution is subleading with respect to $1/(L^d T)$.

Having obtained the correct scaling form  for the marginal model (\eqref{MRGscal}), we can test it numerically.  We performed Monte Carlo (MC) simulations \cite{barkema2001monte} on a three-dimensional cubic lattice with periodic boundary conditions, using the microscopic Hamiltonian \eqref{erboss}, together with the classic Botzmann weight \cite{barkema2001monte}.  We used  lattices with side $L$ ranging from $10$ to $60$ and temperatures $T$ from $10^{-3}$ to $10^{-8}$,  while the parameters of \eqref{erboss} and \eqref{vice} were fixed to $\lambda=J=1$.  We performed standard Metropolis MC with a temperature-dependent Cartesian displacement for the spins (since their length is not fixed) such that the acceptance probability of each move is around $50\%$.  We discard the first $2\times10^5$ MC steps of every simulations, checking every time that we are well above the equilibration time for that specific simulation. The modulus susceptibility is computed via the fluctuation-dissipation relation \cite{Amit1978},
\begin{equation}
	\chi = \frac{1}{TN}\sum_{i,j}\left( \left<|\bm{\sigma_i}||\bm{\sigma_j}|\right>-\left<|\bm{\sigma_i}|\right>\left<|\bm{\sigma_j}|\right>\right), \label{fdtMRG}
\end{equation}
averaging over the MC trajectory.  The soundness of the numerical estimates is checked by using the error analysis presented in \cite{Amit1978}, which makes use of time blocking of data to figure out the adequate simulation length to prevent error underestimation.  We make a small remark for clarity's sake: one might be confused by the fact that in the above equation we have included a  prefactor $1/T$, while we omitted it in the computation of the anomalous finite-size scaling (\eqref{chidim} and following).  This prefactor is harmless in the usual case, but here, since the critical point is $T=0$, it is crucial to get it right.  However, if we look at the definition of the fields we find that there is no inconsistency, since the field of \eqref{chidim} was already rescaled by the square root of $T$ (see passage from \eqref{newbie} to \eqref{erboss}).  Hence, if we compute the susceptibility from the field $\psi$ we do not have to include the prefactor $1/T$ while it must be included when computing it from the original spins $\bm{\sigma}$.

We show in Fig.~\ref{fig:FSS}  the susceptibility for the various system sizes.  Using the scaling variables (right panel), the collapse is quite satisfactory. This result not only strongly supports the theoretical RG calculations, but also confirms that indeed the Landau-Ginzburg Hamiltonian \eqref{Jimi} is the correct effective field theory to describe the modulus mode of the microscopic theory \eqref{erboss}, validating the approximations we made to obtain the field theory.

\section{Conclusions}

The marginal theory has been introduced as a novel form of speed control in highly polarised animal groups, where scale-free correlations of both orientation and speed clash with the standard $\mathrm{O}(n)$ ferromagnetic scenario in the ordered phase, according to which the correlation length of the modulus of the order parameter is finite in the whole symmetry-broken phase. Marginal speed control solves this problem {and it reproduces all the experimental phenomenology \cite{cavagna2022marginal}} by using a bare potential which has zero second derivative with respect to the modulus of the order parameter, thus giving a zero-temperature fixed point. The relative equilibrium field theory has both cubic and quartic vertices, so that a one-loop RG analysis of the critical exponents is non-trivial; moreover, the peculiar nature of the $T=0$ critical point demands that the explicit role of the temperature be treated with care. In the end, the RG flow shows that the critical exponents regulating correlation length and correlation function have the free values $\nu=1/2$ and $\eta=0$. This is supported by the anomalous finite-size scaling of the susceptibility found in Monte Carlo simulations, which confirm that the marginal theory is free for $d=3$.

Assuming that our theoretical results also hold in the off-equilibrium case (which is not certain, despite the weak off-equilibrium effects in starling flocks), one interesting question is whether or not one may observe the free critical exponents in real instances of bird flocks. As a matter of fact, this may be quite tricky, at least with the current type of available data. Previous investigations \cite{cavagna2013boundary} have shown that the ever-changing dynamical inflow of information at the boundary of the flocks may change significantly the bulk decay form of the correlation function, in such a way to screen completely the underlying critical exponents $\nu$ and $\eta$. {Hence the power law decay of the correlation function (which is linked with $\eta$ \cite{Amit1978}) computed in the previous studies of scale-free correlations in starling flocks \cite{cavagna+al_10} is not reproducible with the model we present in this work, which does not take into account dynamical out-of-equilibrium effects on the boundary of the system \cite{cavagna2013boundary}. Moreover, it is not possible to measure independently $\nu$ or $\gamma$ directly from the data \cite{cavagna2022marginal,cavagna2018physics} since it is not clear how to change the temperature (or an equivalent control parameter) of a single flock.} Hence, to test the critical exponents of the marginal model in the wild, one would need a different kind of data, possibly obtained in less perturbed environments than the currently available ones.

From a field-theoretical point of view, it would be interesting to investigate further the co-existence of the zero-temperature critical point, $T=0$, which makes the modulus fluctuations scale-free, and the standard finite critical point, $T=T_c$, where all modes are scale-free. In the symmetry broken phase, the standard transverse correlation length is infinite, due to the Goldstone mode; however, there is a finite length scale in this phase, which regulates the scaling relations below $T_c$, namely the Josephson correlation length, $\xi_J$, which diverges at $T_c$, but {\it decreases} when lowering the temperature below $T_c$ \cite{josephson_relation_1966}. At the same time, the modulus correlation length, $\xi$, {\it increases} in the marginal model when going deeper in the ordered phase. The interplay of these two length scales, which have opposite behaviour in $T$, and their impact on the scaling properties of the theory, remains unclear to us and it is possibly worth of further investigation.

\section*{Acknowledgements}

This work was supported by ERC Advanced Grant RG.BIO (contract n.785932) to ACa.
We thank Irene Giardina for discussions. We also thank V\'ictor Mart\'in Mayor for illuminating advice about finite-size scaling, and in particular for pointing out references \cite{brezin1982investigation} and \cite{brezin1985finite}.
	
\appendix

\section{Mean Field approximation}

\label{app:mean-field}

\subsection{Starting point and general idea}
	
We generalize the theoretical analysis of \cite{cavagna2019CRP} for a model with $n$-component spins $\bm{\sigma}_i$. We want to obtain a closed approximation for the Gibbs free energy \cite{cavagna2019CRP},
\begin{equation}
  g(m) = - J m^2 - \bm{m} \cdot \bm{x}_0 (\bm{m}) - \frac{1}{\beta}\ln \int \mathrm{d} \bm{\sigma} \ e^{-\beta \left[S(\sigma) + \bm{x}_0 (m) \cdot \bm{\sigma} \right]} \label{gbaby}
\end{equation}
where $S(\sigma)=J\sigma^2 + V(\sigma)$ and $\bm{x}_0(\bm{m})$ is an auxiliary variable, defined by the saddle point equation for $N \rightarrow \infty$ \cite{cavagna2019CRP}, which reads,
\begin{equation}
  \bm{m} = \frac{\int \mathrm{d} \bm{\sigma} \ \bm{\sigma} e^{-\beta\left[S(\sigma)+\bm{x}_0\cdot \bm{\sigma}\right]}}{\int \mathrm{d} \bm{\sigma} \ e^{-\beta\left[S(\sigma)+\bm{x}_0\cdot \bm{\sigma}\right]}}. \label{mbaby}
\end{equation}
This equation and the integral in \eqref{gbaby} can be solved numerically for any value of $\beta$ \cite{cavagna2019CRP}, but here we are interested in the asymptotic form of the free energy for large $\beta$ (or $T \rightarrow 0$), hence we perform the integrals in \eqref{gbaby} and \eqref{mbaby}, using once again the saddle point method, this time for $\beta \rightarrow \infty$. Since we want all the corrections up to $O(T^2)$ we have to expand the exponential in each integral up to that order. The saddle point equation for the integrals in $\sigma$ introduces a new player, the saddle point value $\bm{\sigma}_0 (\bm{m})$,
	\begin{equation}
		\nabla S(\bm{\sigma}) \big|_{\bm{\sigma}_0 (\bm{m})} + \bm{x}_0 (\bm{m}) = 0 \label{gradbaby}
	\end{equation}
Now we have to solve \eqref{mbaby}, which will give us an expression for $\bm{\sigma}_0 (\bm{m})$, then use \eqref{gradbaby} to find an expression for $\bm{x}_0(\bm{m})$, and eventually plug everything into \eqref{gbaby}, in order to express the explicit dependence of the Gibbs free-energy on the magnetization (\eqref{gigi}). If we look at \eqref{mbaby} and \eqref{gradbaby}, we can see that $\bm{m}$, $\bm{\sigma}_0$ and $\bm{x}_0$ are parallel (or anti-parallel). Hence it is convenient to write them as $\bm{m}=m\bm{\hat{w}}$, $\bm{\sigma}_0=\sigma_0\bm{\hat{w}}$ and $\bm{x}_0=x_0\bm{\hat{w}}$, where $|\bm{\hat{w}}|=1$. This simplifies our saddle-point calculations, transforming many gradients and Hessian matrices into simple derivatives.
	
\subsection{Computation of necessary terms}
	
We want to expand the Gibbs free-energy \eqref{gbaby} up to $O(T^2)$ (that is $1/\beta^2$).  To accomplish that, we write \eqref{gbaby} expanding the integral in $\bm{\sigma}$, using the saddle point method, which reads,
	\begin{align}
		g(m) &= - J m^2 - x_0 m - \frac{1}{\beta} \ln \left[ \frac{e^{-\beta\left(S_0 + x_0 \sigma_0\right)}}{\det[S_{\alpha \beta}(\sigma_0)]} \left(1 + \frac{1}{\beta}B_0\right)\right] \nonumber \\
		&=-Jm^2 + x_0(\sigma_0 - m) + S_0 +\nonumber\\
		& \ \ \ + \frac{1}{2\beta} \ln \det[S_{\alpha \beta}(\sigma_0)] - \frac{1}{\beta^2} B_0 \label{gteen}
	\end{align}
where $S_0=S(\sigma_0)$, $S_{\alpha \beta}$ is the Hessian matrix of $S$ and $B_0 = B(\sigma_0)$ is the first coefficient of the expansion in $1/\beta$ of the integral in \eqref{gbaby}, which will be computed later on. If we look at \eqref{mbaby} and \eqref{gradbaby}, we can write $\sigma_0$ and $x_0$ as expanding around $m$,
\begin{align}
  &\sigma_0 = m + \frac{1}{\beta} C_m + O(1/\beta^2), \label{szero}\\
  &x_0 = - S'_m - \frac{1}{\beta} S''_m C_m + O(1/\beta^2), \label{xzero}
\end{align}
where $C_m=C(m)$ is the first coefficient of the expansion in $1/\beta$, coming from \eqref{mbaby}, that will be computed later; $S'_m = S'(m)$ and $S''_m = S''(m)$ are respectively the first and second derivative of $S$, from now on this notation will be used for derivatives. If we plug \eqref{szero} and \eqref{xzero} into \eqref{gteen} and keep all terms up to order $1/\beta^2$, we find,
\begin{multline}
  g(m) = V(m) + \frac{1}{\beta} \ln \det \left[S_{\alpha \beta} (m)\right] - \frac{1}{\beta^2} \Bigg[B(m) +  \\
  + \frac{1}{2}S''(m) C^2(m) - \frac{C(m) \left[\det S_{\alpha \beta} (m) \right]'}{2 \det \left[S_{\alpha \beta} (m)\right]}\Bigg] . \label{gadult}
\end{multline}
Even if we want to compute the free energy up to $O(1/\beta^2)$ we do not need to compute the corresponding terms in the expansions \eqref{szero} and \eqref{xzero}, because they cancel out once we substitute them into the free energy.
To compute the term of order $1/\beta$ in \eqref{gadult} we just need to evaluate the determinant of the Hessian of the function $S$ at $m$, the Hessian matrix is diagonal and gives,
\begin{equation}
  \det S_{\alpha \beta} = S''(m) \left[\frac{S'(m)}{m}\right]^{n-1}.
\end{equation}
If we take the logarithm and expand near $m^2 \sim 1$ we find,
\begin{equation}
  \ln \det \left[S_{\alpha \beta} (m)\right] \sim (m^2 - 1)^2,
\end{equation}
which is the term of order $T$ in \eqref{gigi}. Going to next order, we can compute the terms $B(m)$ and $C(m)$ by expanding the integrals of \eqref{gbaby} and \eqref{mbaby} using the saddle point method. After some calculations we find that the leading order in $(m^2 - 1)$, for the term of order $1/\beta^2$ of the \eqref{gadult} is given by the term $B(m)$, which reads,
\begin{multline}
  B(m) \sim - \frac{S_{\alpha \beta \mu \nu}(m)}{24} \left<y_\alpha y_\beta y_\mu y_\nu \right> \sim  - \frac{1}{8} \left[\frac{S''''}{(S'')^2} + ...\right]  \\
       \sim \mathrm{const.} + O(m^2 - 1), \label{bfactor}
\end{multline}
where $S_{\alpha \beta \mu \nu}$ is the fourth order derivatives tensor of $S$ and the $y_\alpha$ are Gaussian distributed variables with,
\begin{align}
  \left<y_\alpha\right> &= 0 \\
  \left<y_\alpha y_\beta\right> &= \left[\delta_{\alpha \beta} \frac{S'(m)}{m} + \frac{m_\alpha m_\beta}{m^2} \left(S''(m) - \frac{S'(m)}{m}\right)\right]^{-1},
\end{align}
therefore we can compute the expected value $\left<y_\alpha y_\beta y_\mu y_\nu \right>$ in \eqref{bfactor} using Wick's theorem \cite{zinnjustin_QFTCF} and the above equation for the covariance. In the end, we obtain that the first non-vanishing term of order $T^2$, apart from the constant, is of order $(m^2 - 1)$, as we can read in \eqref{gigi}.

\section{Initial rescaling of fields}

\label{app:field-rescaling}
	
We now spell out the initial rescaling of fields that links the free energy \eqref{newbie} with the free energy \eqref{Jimi}. If we compute with the mean field approximation the single-particle variance of the spin modulus $s_i=|\bm\sigma_i|$ we obtain,
\begin{align}
  C_0 = \left<s_i^2\right> - \left<s_i\right>^2 \sim T ,
\end{align}
which can also be obtained by computing the connected correlation function in the Gaussian approximation of \eqref{newbie},
\begin{equation}
  \left\langle \varphi(\bm{k})\varphi(\bm{k}')\right\rangle^0_c = \delta(\bm{k}+\bm{k}') \frac{T}{k^2 + a T} \sim \delta(\bm{k}+\bm{k}') \frac{C_0}{k^2 + a T}.
\end{equation}
We see that the $T$ prefactor is problematic, since in the limit of vanishing temperature the correlation function's amplitude $C_0$ vanishes. We want to investigate the regime of small $T$ where the modulus correlation length is large, but we do not want the amplitude of the correlation function itself to vanish. For this reason it seems natural to define a new field,
\begin{equation}
  \psi= \frac{\varphi}{\sqrt{T}} , \label{rescale}
\end{equation}
such that the correlation function of $\psi$ has a fixed amplitude for every temperature,
	\begin{align}
		\left<\psi(\bm{k})\psi(\bm{k}')\right>^0_c = \delta(\bm{k}+\bm{k}') \frac{1}{k^2 + a T}
	\end{align}
at least in the Gaussian and mean-field approximations.  We do not expect great deviations of $C_0$ from the mean field behaviour, given the finding discussed in the main text that the zero-temperature critical point is ruled by the Gaussian fixed point.
	
\section{Renormalization Group calculations}
	
\label{app:RGdetails}

\subsection{Diagrams at one-loop}

The marginal field-theory Hamiltonian \ref{Jimi} has two non-Gaussian vertices: a cubic one with coupling $vT^{3/2}$ ($\blacktriangle$) and a quartic one with coupling $uT$ ($\bullet$) (see Fig. \ref{fig:vertices}).  We can combine these two vertices, in order to form all the possible one-loop diagrams with an arbitrary number of external legs. Since we evaluate the renormalized couplings only up to the term $\psi^4$, we stop at four external legs. All the diagrams with more than four external legs give a correction to higher order terms that we do not include in \eqref{Jimi} because they are RG-irrelevant.  The diagrams that give a contribution to the renormalization of temperature $T$ are,
\begin{align}
  &\raisebox{-.5\height}{\includegraphics[width=0.15 \linewidth]{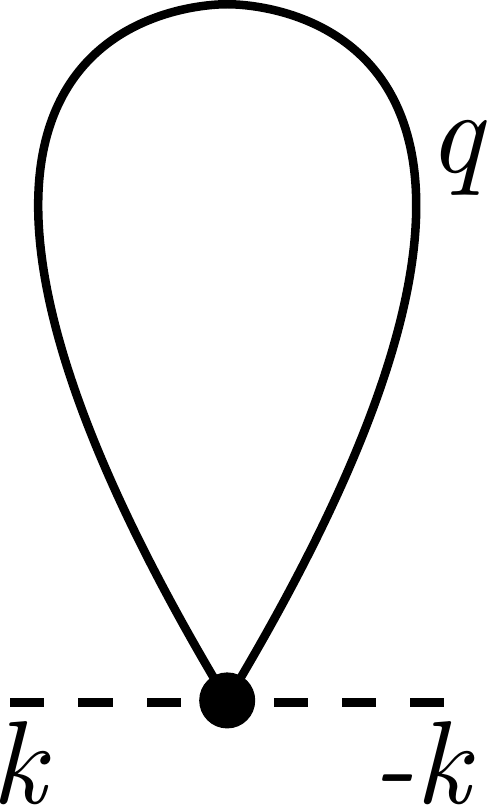}}
    \ \ \ \ \ \ \ \ \ \ \ \ \ \ \ = 12 uT \int\limits_{\Lambda/b}^{\Lambda} \frac{\mathrm{d}^d q}{(2 \pi)^d} \frac{1}{q^2 + T} \nonumber \\
  &\raisebox{-.5\height}{\includegraphics[width=0.4 \linewidth]{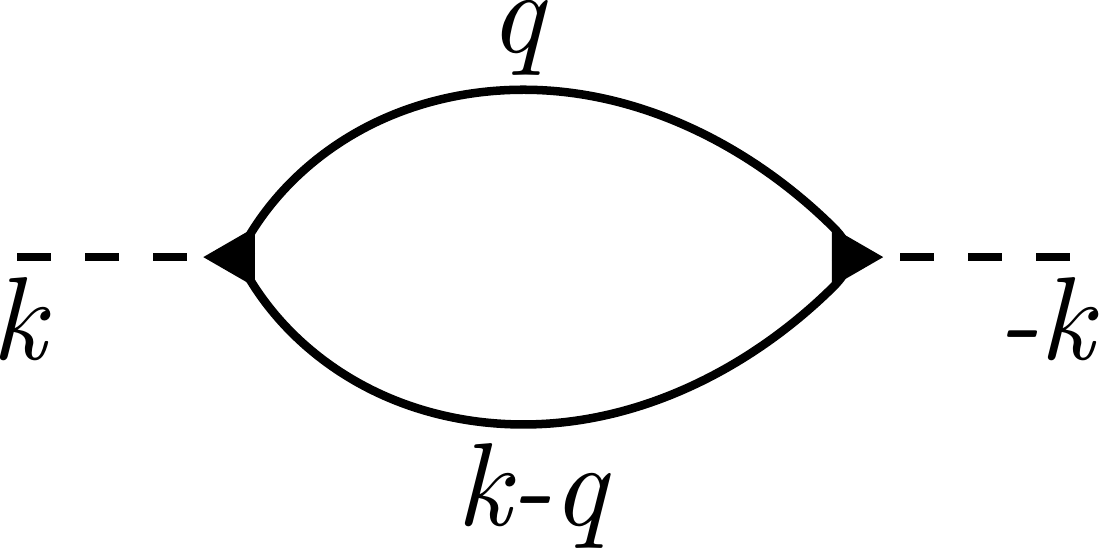}}
  \\ &= -18 v^2T^3 \int\limits_{\Lambda/b}^{\Lambda} \frac{\mathrm{d}^d q}{(2 \pi)^d} \frac{1}{\left(q^2 + T\right)\left[(\bm{k}-\bm{q})^2 + T\right]} \label{twoleg}
\end{align}

The renormalization of $vT^{3/2}$ comes from
\begin{align}
  &\raisebox{-.5\height}{\includegraphics[width=0.35\linewidth]{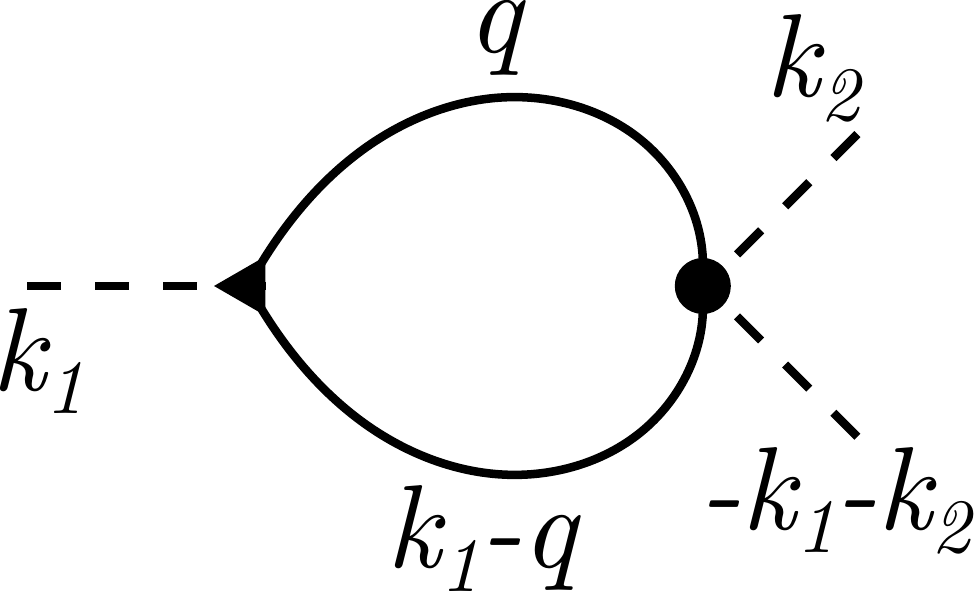}}\nonumber \\
  &= -36 uvT^{5/2} \int\limits_{\Lambda/b}^{\Lambda} \frac{\mathrm{d}^d q}{(2 \pi)^d} \frac{1}{\left(q^2 + T\right)\left[(\bm{k}_1-\bm{q})^2 + T\right]}
\end{align}
\begin{align}
  &\raisebox{-.5\height}{\includegraphics[width=0.35\linewidth]{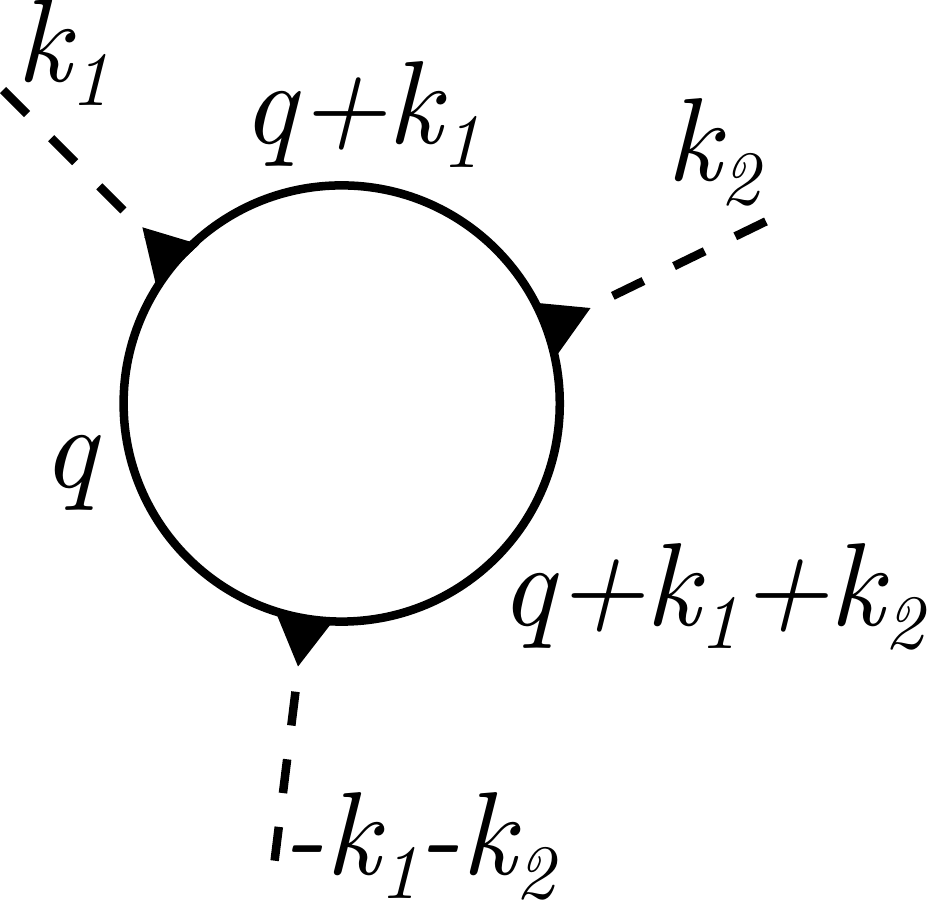}} \nonumber \\
  &= 36 v^3T^{9/2} \int\limits_{\Lambda/b}^{\Lambda} \frac{\mathrm{d}^d q}{(2 \pi)^d} \frac{1}{\left(q^2 + T\right)\left[(\bm{k}_1+\bm{q})^2 + T\right]} \times \nonumber \\
  &\times	\frac{1}{\left[(\bm{k}_1+\bm{k}_2+\bm{q})^2 + T\right]}
\end{align}

finally the renormalization of $uT$ is due to
\begin{align}
  &\raisebox{-.5\height}{\includegraphics[width=0.4\linewidth]{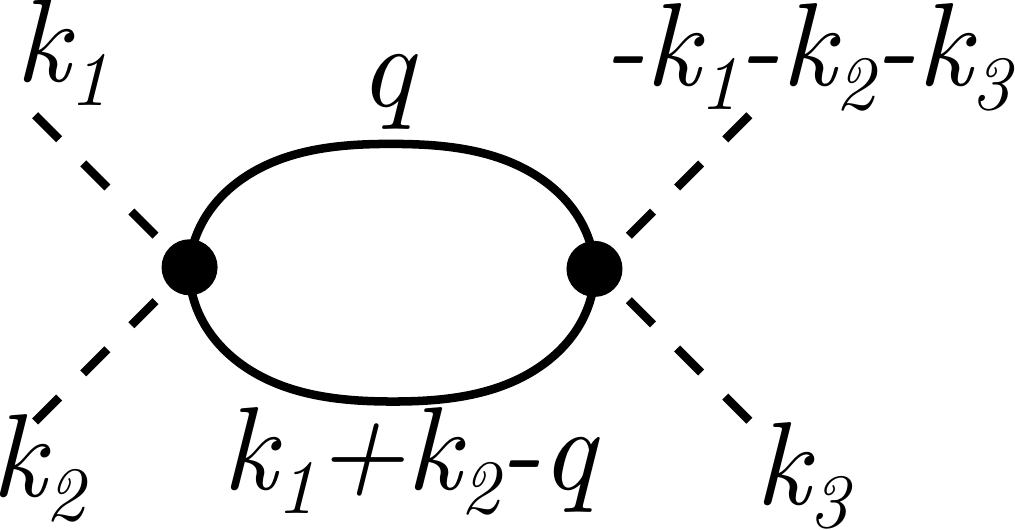}} \nonumber \\
  &=-36 u^2T^2 \int\limits_{\Lambda/b}^{\Lambda} \frac{\mathrm{d}^d q}{(2 \pi)^d} \frac{1}{\left(q^2 + T\right)\left[(\bm{k}_1+\bm{k}_2-\bm{q})^2 + T\right]} \\
  &\raisebox{-.5\height}{\includegraphics[width=0.4\linewidth]{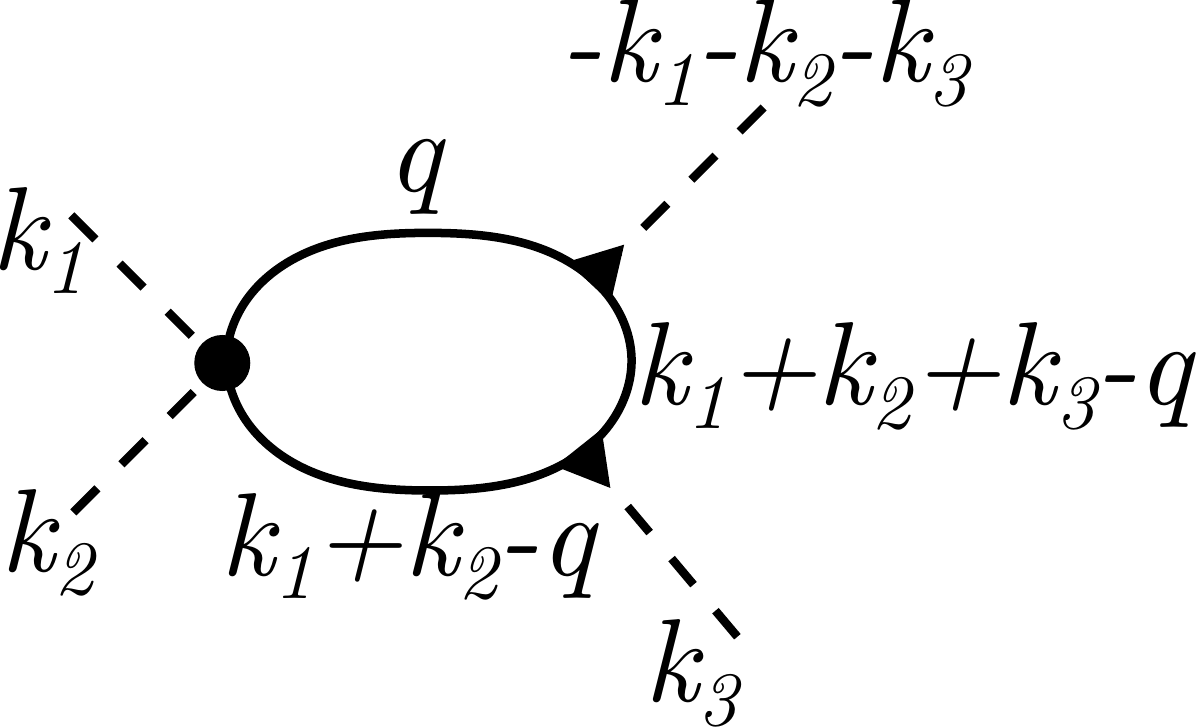}} \nonumber \\
  &= 216 uv^2T^4 \int\limits_{\Lambda/b}^{\Lambda} \frac{\mathrm{d}^d q}{(2 \pi)^d} \frac{1}{\left(q^2 + T\right)\left[(\bm{k}_1+\bm{k}_2-\bm{q})^2 + T\right]} \times \nonumber \\ &\times \frac{1}{\left[(\bm{k}_1+\bm{k}_2+\bm{k}_3-\bm{q})^2 + T\right]}
  \end{align}
  \begin{align}
  &\raisebox{-.5\height}{\includegraphics[width=0.4\linewidth]{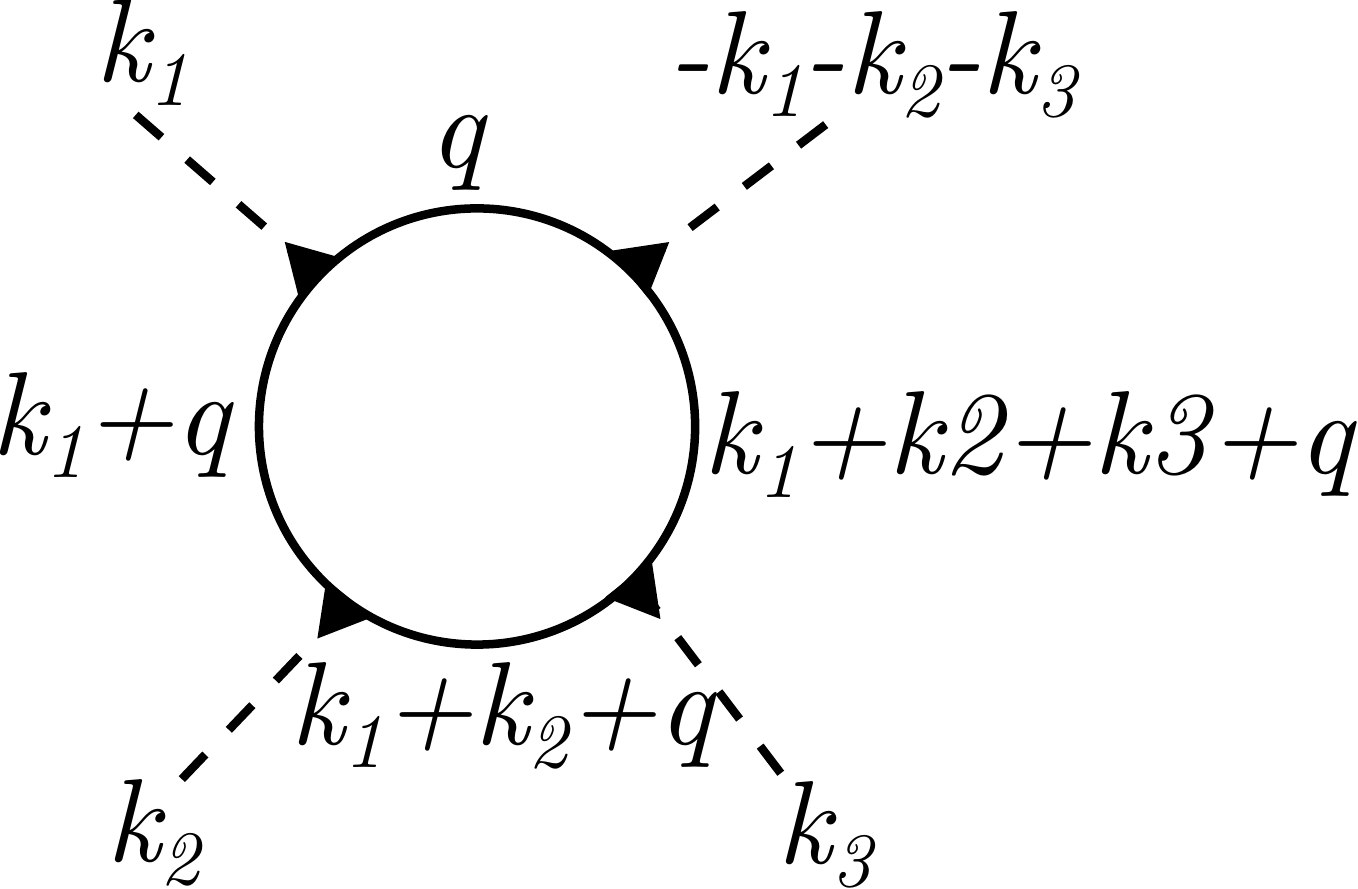}} \nonumber \\
  &= -162 v^4T^6 \int\limits_{\Lambda/b}^{\Lambda} \frac{\mathrm{d}^d q}{(2 \pi)^d} \frac{1}{\left(q^2 + T\right)\left[(\bm{k}_1+\bm{q})^2 + T\right]}\times \nonumber \\
  &\times \frac{1}{\left[(\bm{k}_1+\bm{k}_2+\bm{q})^2 + T\right]\left[(\bm{k}_1+\bm{k}_2+\bm{k}_3+\bm{q})^2+T\right]}
	\end{align}
Dashed lines represent fields with momentum $k<\Lambda/b$ (off-shell), while solid lines represent integrated fields with momentum $\Lambda/b < k < \Lambda$  (on-shell). Since we are interested in the corrections to the couplings of momentum-independent terms ($\psi^2$, $\psi^3$ and $\psi^4$) we can compute all these diagrams at zero external momentum and obtain the corrections of Eqs.~\ref{athos}, \ref{porthos} and~\ref{aramis}.

\subsection{The linear term}
\label{sec:linear-term}
We have ignored in the  Landau-Ginzburg free energy \eqref{Jimi} a linear term in $\psi$ that would have read $c T^{3/2} \psi$ (following the mean-field Gibbs free energy \eqref{latrottola} and using the rescaling $\psi=\varphi/\sqrt{T}$), where $c$ is a constant independent of temperature.  We made this choice because the linear term can be removed with a simple shift of the field by a constant value. If we include the linear term in the theory, we find that the packed constant $cT^{3/2}$ is corrected by the diagram,
\begin{align}
  \raisebox{-.5\height}{\includegraphics[width=0.35\linewidth]{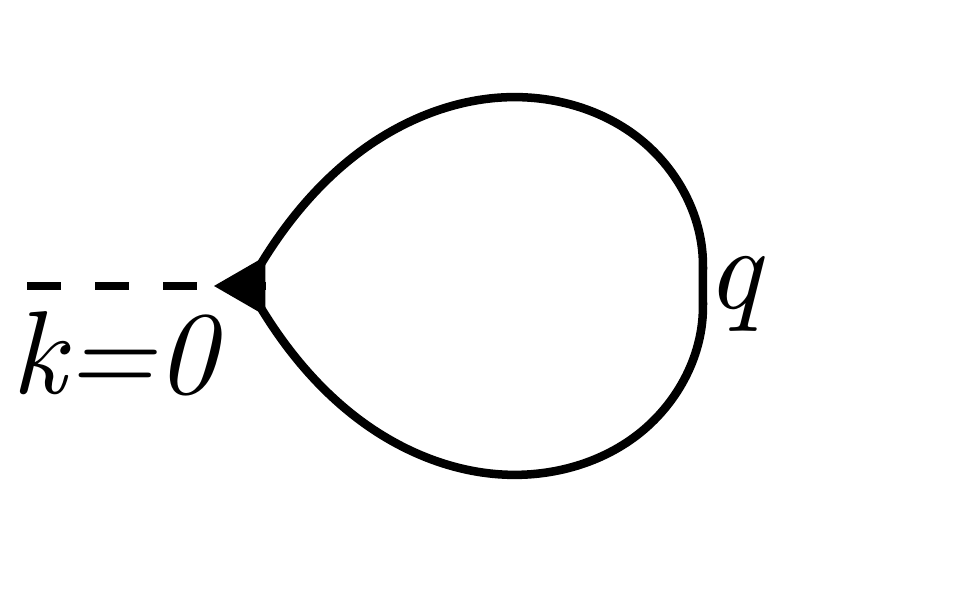}}\ \ \ \ \ \ \ \ = 3 vT^{3/2} \int\limits_{\Lambda/b}^{\Lambda} \frac{\mathrm{d}^d q}{(2 \pi)^d} \frac{1}{\left(q^2 + T\right)}
\end{align}
After the same calculations that we did for the other terms, the $\beta$ function of the parameter $c$ is,
\begin{equation}
	\beta_c = -\frac{c}{2} + 3v\Lambda^{d-2} - 18 uc\Lambda^{d-2},
\end{equation}
which tells us that at the Gaussian fixed point $v^*=u^*=c^*=0$, the parameter $c$ is also irrelevant.  We  also note that the linear term does not produce any  diagram which could contribute to the renormalization of the other couplings, \eqref{eq:renormhat}.  Hence the phenomenology that we have described in the main text does not change, even if we add the linear term.  Also in this case the differential equation $\dv{c}{x}=\beta_c(c,v,u)$ is exactly solvable and it reads (for $d=3$),
\begin{equation}
  c(x)=\frac{c_0 e^{-x/2} +3\Lambda v_0 e^{-x/2}(1-e^{-x})}{\left[1+12\Lambda u_0(1-e^{-x})\right]^{3/2}},
\end{equation}
which means that for any starting condition the parameter $c$ flows to $0$.

\subsection{Higher order couplings}

To check for the relevance of terms of order higher than $\psi^4$ we need to know the naive scaling dimension of their $T$-independent couplings.  To see that, go back to \eqref{latrottola}, which gives the dependence on $T$ of each coupling, based on the mean-field Gibbs free energy. We find that, before the rescaling $\psi=\varphi/\sqrt{T}$, the higher order terms can be written as,
\begin{multline}
	\mathcal{H}_{high} = \frac{1}{T} \int \mathrm{d}^d x \ \Bigg\{u_5 \varphi^5 + u_6 \varphi^6 + \dots + u_8 \varphi^8 +  \\
 +u_9 T \varphi^9 + u_{10} T \varphi^{10} + \dots + u_n T \varphi^n + \dots\Bigg\},
\end{multline}
where every $u_l$ is a constant independent of $T$. We find this dependence on $T$ from \eqref{gadult}, where we can see that the lowest order (in $T$) that generates the terms from $(m-1)^5$ up to $(m-1)^8$ is the first term (the bare marginal potential) hence their couplings do not depend on $T$. On the other hand, the lowest order term that generates powers from $(m-1)^9$ and above is the logarithm of order $T$.  Upon rescaling the field we have,
\begin{multline}
  \mathcal{H}_{high} = \int \mathrm{d}^d x \ \Bigg\{u_5 T^{3/2} \psi^5 + \dots + u_8 T^3 \psi^8 + \\
	+ u_9 T^{9/2} \psi^9 + \dots + u_n T^{n/2} \psi^n + \dots\Bigg\},
\end{multline}
which can be expressed as,
\begin{equation}
  u_n \psi^n \rightarrow
  \begin{cases}
    T^{n/2-1} \ \ \ \ \ \ \ \ \ \ \text{for} \ 4<n<9\\
		T^{n/2} \ \ \ \ \ \ \ \ \ \ \ \ \ \text{for} \ n \geq 9
              \end{cases}
\end{equation}
Using the expressions above we can compute the naive scaling dimensions of the $u_n$ couplings, which are
\begin{equation}
	\left[u_n\right] = \begin{cases}
		2+d\left(1-\frac{n}{2}\right) \ \ \ \ \ \ \ \ \ \ \text{for} \ 4<n<9\\
		d\left(1-\frac{n}{2}\right)\ \ \ \ \ \ \ \ \ \ \ \ \ \ \ \text{for} \ n \geq 9
	\end{cases}
\end{equation}
For $d=3$ we see that $[u_n]<0$ for all $n>4$, hence the Gaussian fixed point $v^*=u^*=u_n^*=0$, remains stable even after adding higher-order terms.

\end{document}